\documentclass[amsmath,amssymb,superscriptaddress,showpacs,nofootinbib]{revtex4-1}

\usepackage{bm,amsmath,amsthm,amssymb,hyperref,graphicx,float,color}
\usepackage{changes}

\newcommand{\be}{\begin{equation}}
\newcommand{\ee}{\end{equation}}
\newcommand{\bea}{\begin{eqnarray}}
\newcommand{\eea}{\end{eqnarray}}
\newcommand{\6}{\partial}

\newcommand{\inti}{\int_{-\infty}^{+\infty}}

\newcommand{\lam}{\lambda}

\newcommand{\la}{v}

\newcommand{\g}{\gamma}

\newcommand{\au}{\mathfrak{a}_1}
\newcommand{\ad}{\mathfrak{a}_2}

\newcommand{\R}{\textsf{R}}

\newcommand{\tr}{\textsf{t}}

\begin{document}

\title{Thermodynamics, contact and density profiles of the repulsive Gaudin-Yang model}
\author{Ovidiu I. P\^a\c tu  }
\affiliation{Institute for Space Sciences, Bucharest-M\u{a}gurele, R 077125, Romania}
\author{Andreas Kl\"umper}
\affiliation{Fachbereich C – Physik, Bergische Universit\"at  Wuppertal, 42097 Wuppertal, Germany}


\begin{abstract}

We address the problem of computing the thermodynamic properties of the
repulsive one-dimensional two-component Fermi gas with contact interaction,
also known as the Gaudin-Yang model. Using a specific lattice embedding and
the quantum transfer matrix we derive an exact system of only two nonlinear
integral equations for the thermodynamics of the homogeneous model which is
valid for all temperatures and values of the chemical potential, magnetic
field and coupling strength. This system allows for an easy and extremely
accurate calculation of thermodynamic properties circumventing the
difficulties associated with the truncation of the thermodynamic Bethe ansatz
system of equations. We present extensive results for the densities,
polarization, magnetic susceptibility, specific heat, interaction energy, Tan
contact and local correlation function of opposite spins.  Our results show
that at low and intermediate temperatures the experimentally accessible
contact is a nonmonotonic function of the coupling strength. As a function of
the temperature the contact presents a pronounced local minimum in the
Tonks-Girardeau regime which signals an abrupt change of the momentum
distribution in a small interval of temperature.  The density profiles of the
system in the presence of a harmonic trapping potential are computed using the
exact solution of the homogeneous model coupled with the local density
approximation. We find that at finite temperature the density profile presents
a double shell structure (partially polarized center and fully polarized
wings) only when the polarization in the center of the trap is above a
critical value which is monotonically increasing with temperature.

\end{abstract}

\maketitle

\section{Introduction}

The Gaudin-Yang model \cite{Ga1,Y1} which describes one-dimensional
spin-$\frac{1}{2}$ fermions interacting via a delta-function potential has a
long and distinguished history being the subject of theoretical investigations
for more than fifty years \cite{GBL}. This experimentally realized model
\cite{MSGKE,LRPLP} not only  presents an incredibly rich physics
of which we mention Tomonaga-Luttinger \cite{Gia} and incoherent spin
Luttinger liquids \cite{BL,Ber,ISLL1,ISLL2}, Bardeen-Cooper-Schrieffer
\cite{BCS1,BCS2} and Fulde-Ferrel-Larkin-Ovchinnikov like pairing, spin-charge
separation, Fermi polarons \cite{McGuire} and quantum criticality and scaling
\cite{QC}, but is also amenable to an exact solution which allows for a
parameter free comparison of theory and experiment. From the historical point
of view the Gaudin-Yang model, also known as the two-component Fermi gas
(2CFG), was the first multi-component system solved by the nested Bethe ansatz
(BA) (the particular cases of one and two particles with spin up in a sea of
opposite spins were considered in \cite{McGuire,FL1}). The thermodynamics of
the 2CFG in the framework of the thermodynamic Bethe ansatz (TBA) was derived
independently by Takahashi and Lai \cite{Tak1,Tbook,Lai}. The experimental
advances of the last 20 years in the field of ultracold atomic gases which
allowed for the creation and manipulation of 1D multicomponent systems renewed
the interest in such models which were investigated further using a large
variety of exact and approximate methods
\cite{Tokatly,Astrak1,Fuchs,IidaW,Orso,HLD,ColTat,LGSB}.

At zero temperature the groundstate in the thermodynamic limit of the
Gaudin-Yang model is characterized by a system of two Fredholm integral
equations whose solution allows to derive the phase diagram
\cite{Orso,HLD,ColTat}.  The situation at finite temperature is much more
complicated. This is due to the fact that the application of TBA produces an
infinite system of nonlinear integral equations which are very difficult to
investigate even numerically. This is the main reason why the vast majority of
results regarding the thermodynamic behaviour of the 2CFG found in the
literature are restricted to a small region of the relevant parameters
(temperature, chemical potential, magnetic field and coupling strength). Even
though numerical schemes to treat the TBA equations have been developed
\cite{TS,CKB,KC,CJGZ} the need for an efficient thermodynamic description of
multicomponent 1D systems cannot be overstated.

In this paper we address this problem by deriving a system of only two
integral equations characterizing the thermodynamics of the repulsive
Gaudin-Yang model which is valid for all values of the physical parameters and
from which physical information can be easily extracted numerically. We employ
the same method we have used in the case of the two-component Bose gas
\cite{KP1,KP2} which utilizes the fact that the 2CFG can be obtained as the
scaling limit of the Perk-Schultz spin chain with a specific grading
\cite{PS1,Schul1,BVV,dV,dVL,L} and the quantum transfer matrix
\cite{MS,Koma,SI1,K1,K2}.  The largest eigenvalue of the quantum transfer
matrix gives the free energy of the associated lattice model which means that
performing the same scaling limit we obtain the grandcanonical potential of
the continuum model. In the homogeneous case we present extensive results for
the densities, polarization, susceptibility, specific heat, Tan contact and
local correlation function of opposite spins for a wide range of coupling
strengths, temperatures and chemical potentials.  We find that, for $T>0$ and
any value of the polarization, the contact \cite{Tan1,OD,BP1,ZL,WTC,BZ,VZM,WC}
is a nonmonotonic function of both coupling strength and temperature. The
implication of this unusual behavior can be more easily understood if we take
into account that for delta-function interactions the contact $C$ controls the
tail of the momentum distribution via $\tilde n(k)\sim C/k^4$ with
$k\rightarrow\infty$.  The local maxima or minima of the contact result in
abrupt changes of $\tilde n(k)$, which can be experimentally detected, and can
be used for accurate thermometry. The reconstruction of the momentum
distribution as a function of the temperature for the balanced impenetrable
system was first described by Cheianov, Smith and Zvonarev in \cite{CSZ}.

The experimentally relevant situation in which the system is subjected to an
external harmonical potential is treated using the local density approximation
\cite{KGDS2, Orso,HLD,ColTat} and the solution of the homogeneous
system. Compared with the zero temperature case \cite{ColTat}, at
low-temperatures the density profiles present a double shell structure with
partially polarized center and polarized wings only if the polarization at the
center of the trap is above a critical value which depends on the
temperature. As we increase the temperature eventually the entire system
becomes partially polarized.

The plan of the paper is as follows. In Section \ref{s2} we introduce the
Gaudin-Yang model and present the zero and finite temperature Bethe ansatz
solution. The new thermodynamic description of the model is introduced in
Section \ref{s3}.  Numerical data for the densities, polarization,
susceptibility, specific heat and Tan contact are reported in Sections
\ref{s4} and \ref{s5}. The density profiles are investigated in Section
\ref{s6}. The derivation of the NLIEs from the similar result for the
Perk-Schultz spin chain is performed in Sections \ref{s7}, \ref{s8} and
\ref{s9}.  The solution of the generalized Perk-Schultz model can be found in
Appendix \ref{a1}.

\section{The Gaudin-Yang model}\label{s2}

The Gaudin-Yang model \cite{Ga1,Y1} describes one-dimensional
spin-$\frac{1}{2}$ fermions interacting via a delta-function potential and
represents the natural extension in the fermionic case of the Lieb-Liniger
model \cite{LL}. In  second quantization the Hamiltonian reads
\begin{align}\label{Hc}
\mathcal{H}_{GY}=\int_0^{L_F} dx \left[\frac{\hbar^2}{2m} (\6_x \Psi^\dagger \6_x \Psi)
+\frac{g}{2}\, :(\Psi^\dagger \Psi)^2:+(V(x)-\mu)(\Psi^\dagger \Psi)-H(\Psi^\dagger\sigma_z\Psi)\right]\, ,
\end{align}
where
$
\Psi=\left(\begin{array}{c} \Psi_\uparrow(x)\\ \Psi_\downarrow(x)\end{array}\right)\, ,\
\Psi^\dagger=\left(\begin{array}{cc} \Psi_\uparrow^\dagger(x) & \Psi_\downarrow^\dagger(x)\end{array}\right)\, ,\
\sigma_z=\left(\begin{array}{cc}1&0\\ 0&-1\end{array}\right)\, , \
$
$V(x)$ is the trapping potential, $\mu$ is the chemical potential, $H$ the
magnetic field and we consider periodic boundary conditions on a segment of
length $L_F$. $\Psi_{\uparrow,\downarrow}(x)$ are fermionic fields which
satisfy canonical anticommutation relations $\{ \Psi_a(x),\Psi_b^\dagger(y)\}
=\delta_{a,b}\delta(x-y)\, .$ In (\ref{Hc}) the symbol $:\  :$ denotes
normal ordering.  The coupling constant $g=\hbar^2 c/m $ can be
positive or negative corresponding to
repulsive or attractive interactions. In this paper we will consider only
the repulsive case $g>0.$ Due to the spin-independent interaction the Zeeman
term in the Hamiltonian (\ref{Hc}) is a conserved quantity. In the following
it will be preferable to use units of $\hbar=2m=1$ and introduce the
dimensionless coupling constant $\gamma=c/n=-2/(a_{1D} n)$ where $n$ is the
density of the system and $a_{1D}$ the scattering length. In terms of this
coupling constant the strong coupling regime is defined as $\gamma\gg 1$ and
the weak coupling regime as $\gamma\ll 1$.

The Gaudin-Yang model is exactly solvable only when $V(x)=0$. However, in the
case of a sufficiently shallow trap, the system can be considered locally
homogeneous and the local density approximation (LDA) \cite{Orso,HLD,
  ColTat,KGDS2} can be used. The LDA coupled with the solution of the
homogeneous system allows for accurate predictions of the density profiles
which will be computed in Section \ref{s6}. Therefore, we will first present
the solution in the $V(x)=0$ case. For a system with $M$ particles of which
$M_\uparrow$ have spin up and $M_\downarrow$ have spin down the energy
spectrum of the homogeneous system is \cite{Ga1,Y1}
\be\label{ef}
E_{GY}=\sum_{j=1}^M \bar e_0(k^{(1)}_j)-H(M_\uparrow-M_\downarrow)\, ,\ \ \ \ \bar e_0(k)=k^2-\mu\, ,
\ee
where the charge and spin rapidities $\{k^{(1)}_j\}\, , \{k^{(2)}_j\}$ satisfy
the Bethe Ansatz equations (BAEs)
\begin{subequations}\label{BEc}
\begin{align}
&e^{ik^{(1)}_sL_F}=
\prod_{p=1}^{M_\downarrow}\frac{k^{(1)}_s-k^{(2)}_p+ic/2}{k^{(1)}_s-k^{(2)}_p-ic/2}\, ,\ \  s=1,\cdots,M\, , \\
&\prod_{j=1}^M\frac{k^{(2)}_l-k^{(1)}_j+ic/2}{k^{(2)}_l-k^{(1)}_j-ic/2}=
\prod_{\substack{p=1\\p\ne l}}^{M_\downarrow}\frac{k^{(2)}_l-k^{(2)}_p+ic}{k^{(2)}_l-k^{(2)}_p-ic}\, , \ \ l=1,\cdots,M_\downarrow\, .
\end{align}
\end{subequations}
The solution at finite temperature is a very difficult task to accomplish even
though the system is integrable.  Assuming the string hypothesis, Takahashi and
Lai \cite{Tak1,Tbook,Lai} obtained for the grandcanonical potential
($\beta=1/T$ with $T$ the temperature)
$\phi_{TBA}(\beta,\mu,H)=-\frac{1}{2\pi\beta}\inti dk\, \ln(1+\zeta(k))\, $
with $\zeta(k)$ satisfying the following infinite system of nonlinear integral
equations (NLIEs)
\begin{subequations}\label{TBAeq}
\begin{align}
&\ln \zeta(k)=-\beta(k^2-\mu)+R\ast\ln(1+\zeta(k))+f\ast\ln(1+\eta_1(k))\, ,\\
&\ln \eta_1(k)=f\ast\left(\ln(1+\eta_{2}(k))-\ln(1+\zeta(k))\right)\, ,\\
&\ln \eta_n(k)=f\ast\left(\ln(1+\eta_{n-1}(k))+\ln(1+\eta_{n+1}(k))\right)\, ,\ \ n=2,\cdots,\infty\, ,
\end{align}
\end{subequations}
together with the asymptotic condition $\lim_{n\rightarrow\infty} \ln
\eta_n(k)/n=2\beta H$. In Eqs.~(\ref{TBAeq}), which are also known as the
thermodynamic Bethe ansatz equations, $g*h(k)\equiv \inti g(k-k')h(k')\ dk'$,
$f(k)=1/[2c\cosh(\pi k/c)], $ $R(k)=b_1*f(k)$ with
$b_1(k)=c/[2\pi((c/2)^2+k^2)]$. It is clear that extracting physically
relevant information from the TBA equations is very hard even from the
numerical point of view. In general, numerical implementations require the
truncation of the equations after a certain level, an approximation which
introduces uncontrollable errors especially in the intermediate and
high-temperature regime.

\section{Efficient thermodynamic description of the repulsive Gaudin-Yang model}\label{s3}

A similar situation is encountered in the case of the one-dimensional
two-component repulsive Bose gas (2CBG).  Using a lattice embedding and the
quantum transfer matrix in \cite{KP1,KP2} we derived a simple system of two
NLIEs characterizing the thermodynamics of the 2CBG which circumvents the
problems associated with the TBA equations. The same method can be applied in
the case of the Gaudin-Yang model (the derivation is presented in Sections
\ref{s8} and \ref{s9}) obtaining for the grandcanonical potential per length
\begin{figure}
\includegraphics[width=0.6\linewidth]{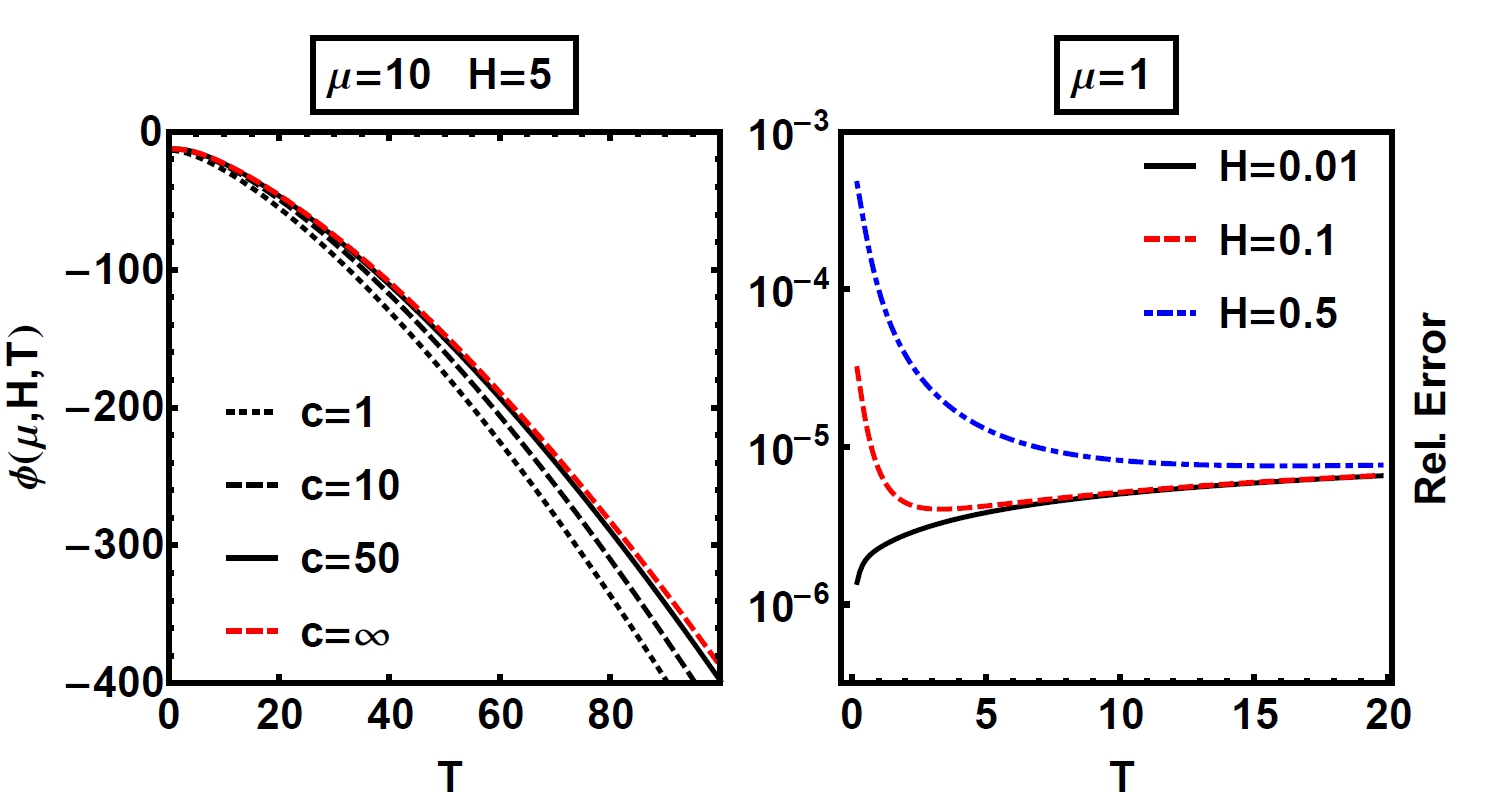}
\caption{(Color online) Left panel. Temperature dependence of the
  grandcanonical potential calculated using Eq.~(\ref{Ggas}) for $\mu=10\,
  ,H=5$ and $c=1$ (dotted black line), $c=10$ (dashed black line) and $c=50$
  (black line). In the limit of impenetrable particles we reproduce
  Takahashi's result (\ref{impf}) (dashed red line). Right panel.  Relative
  error $|\phi-\phi_{TBA}|/|\phi|$ with $\phi$ the grandcanonical potential
  given by Eq.~(\ref{Ggas}) and $\phi_{TBA}$ calculated from the TBA equations
  (\ref{TBAeq}) for $c=1,\mu=1$ and magnetic fields $H=\{0.01,0.1,0.5\}$.
  Grandcanonical potential per length and  temperature in units of $\phi_0$ and $T_0$ \cite{uni}.
  }

\label{Im}
\end{figure}
\be\label{Ggas}
\phi(\beta,\mu,H)=-\frac{1}{2\pi\beta}\int_{\mathbb{R}}\ln(1+a_1(k))+\ln(1+a_2(k))\, dk\, ,
\ee
where $a_{1,2}(k)$ are auxiliary functions satisfying ($\varepsilon\rightarrow 0$)
\begin{subequations}\label{NLIEgas}
\begin{align}
\ln a_1(k)&=-\beta(k^2-\mu-H)
+\int_{\mathbb{R}} K_2^F(k-k'-i\varepsilon)\ln(1+a_2(k'))\, dk'\, ,\\
\ln a_2(k)&=-\beta(k^2-\mu+H)+\int_{\mathbb{R}} K_1^F(k-k'+i\varepsilon)\ln(1+a_1(k'))\, dk'
\, ,
\end{align}
\end{subequations}
and
\be
 K_1^F(k)=\frac{1}{2\pi }\frac{c}{k(k+ic)}\, ,\ \
 K_2^F(k)=\frac{1}{2\pi }\frac{c}{k(k-ic)}\, .
\ee
We would like to stress that this system of equation is exact and valid for
all values of chemical potential, magnetic fields (including the balanced case
$H=0$), positive coupling strengths and temperature.  We also note that
(\ref{NLIEgas}) differs from the result for the
2CBG derived in \cite{KP1} by dropping the diagonal terms.

In the left panel of Fig. \ref{Im} we present results for the temperature
dependence of the grandcanonical potential of a system with $\mu=10$ and $H=5$
showing how for increasing values of the coupling strength $c=\{1,10,50\}$ we
approach Takahashi's result for impenetrable particles \cite{Tak1,Tbook}
\be\label{impf}
\phi_\infty(\beta,\mu,H)=-\frac{1}{2\pi\beta}\int_{\mathbb{R}}dk\,\ln\left(1+2\cosh(\beta
H)e^{-\beta(k^2-\mu)}\right)\, .  \ee A comparison between the grandcanonical
potential for a system with $c=1$ and $\mu=1$ computed using Eq.~(\ref{Ggas})
and the TBA equations (\ref{TBAeq}) truncated at the $n=30$ level can be found
in the right panel of Fig. \ref{Im}.  The results show almost perfect
agreement which not only confirms the validity of our results but also
supports the string hypothesis in the case of the repulsive Gaudin-Yang model.

In the noninteracting limit, $c\rightarrow 0$, using $\lim_{c\rightarrow 0}
K_2^F(k-i\varepsilon)$ $=\lim_{c\rightarrow 0} K_1^F(k+i\varepsilon)=0$ it is
easy to show that (\ref{Ggas}) reduces to
$
\phi_0(\beta,\mu,H)=\frac{1}{2\pi\beta}\int_{\mathbb{R}}
\ln(1+e^{-\beta(k^2-\mu-H)})+\ln(1+e^{-\beta(k^2-\mu+H)})dk\, ,
$
which, as expected, is the result for free fermions in a magnetic field.

\section{Thermodynamic properties  of the homogeneous system}\label{s4}

Despite being the first multi-component model solved by Bethe ansatz at zero
and finite temperature our knowledge of the thermodynamic properties of the
Gaudin-Yang model is still very limited. Fueled by the interest in the FFLO
pairing phase the case of attractive interactions has received more attention
\cite{Orso,HLD,HJLPAD,BAD,GYFBLL} but even in this case a complete characterization
of the thermodynamical properties is still lacking.
The situation is even more dire in the repulsive case. The phase diagram at $T=0$
was derived by Colom\'e-Tatch\'e \cite{ColTat} and the low-temperature
strong-coupling regime was investigated by Lee, Guan, Sakai and Batchelor in
\cite{LGSB}.
\begin{figure}
\includegraphics[width=0.9\linewidth]{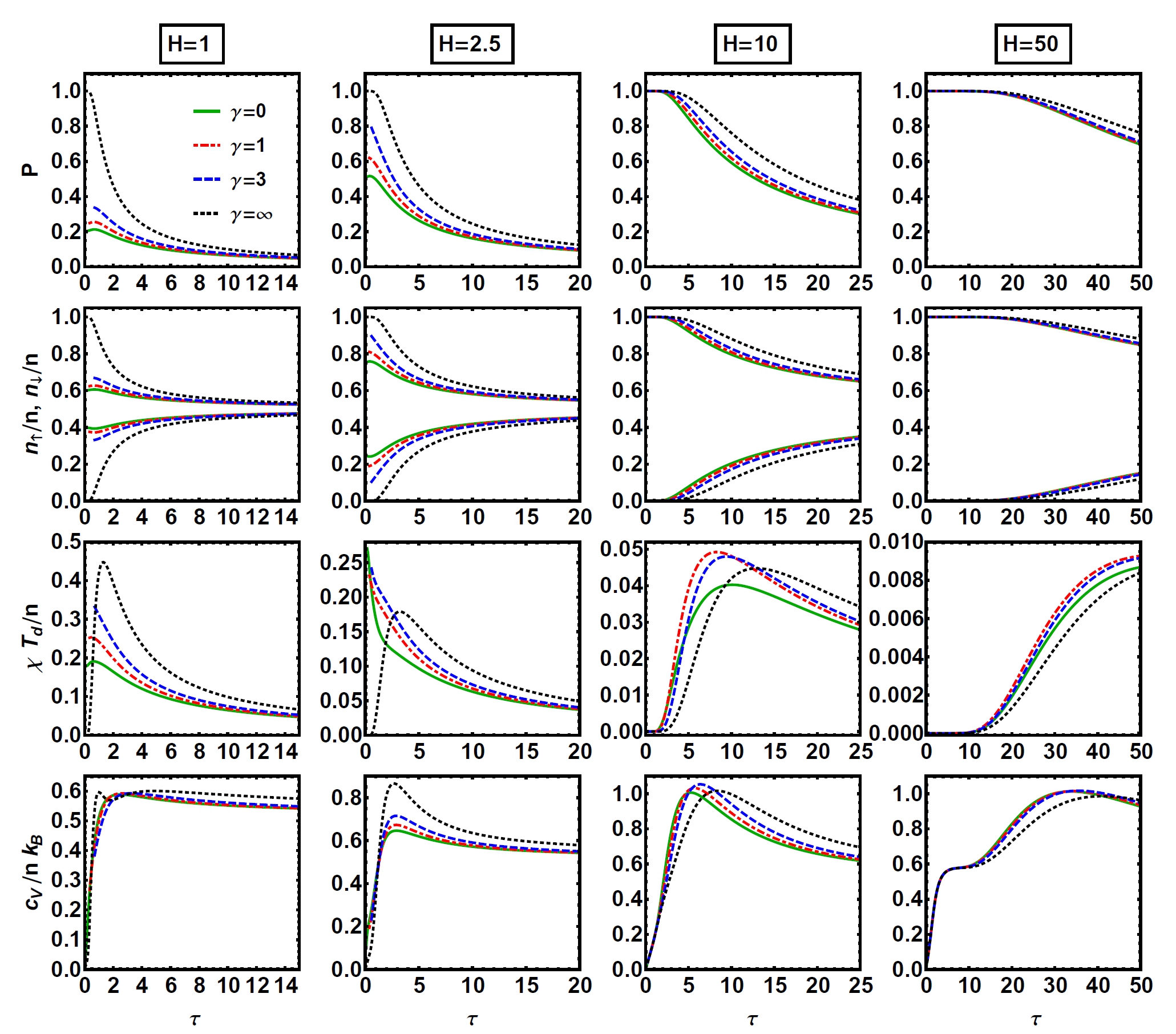}
\caption{(Color online) Plots of the polarization (first row), densities of
spin up and down particles (second row), susceptibility per length normalized by $n/T_d$ (third row)
and specific heat per length normalized by $n k_B$ (bottom
row) as functions of the reduced temperature $\tau=T/T_d=T/n^2$ for four values of
the dimensionless coupling strength $\g=\{0,1,3,\infty\}$ and magnetic field
$H=\{1,2.5,10,50\}$. The total density is fixed to $n=1$.  }
\label{TData}
\end{figure}
The main reason behind this information scarcity is the complexity of the TBA
equations and other methods used in the investigation of thermodynamic
properties, like lattice Monte-Carlo calculations, which require rather
cumbersome numerical schemes in order to obtain accurate numerical data. In
contrast, our equations (\ref{NLIEgas}), can be accurately and easily solved
employing a simple iterative procedure with the convolutions treated using the
Fast Fourier Transform and the convolution theorem. This scheme allows us to
probe a wide region of the parameter space with the exception of the low
polarization at low-temperatures regime ($H\ll \mu$ and $T\rightarrow
0$). This regime requires a more careful treatment of the NLIEs which will be
deferred to a future publication.

In Fig.~\ref{TData} we present for several values of the magnetic field the
dependence on the reduced temperature $\tau=T/T_d=T/n^2$ of the densities
$n_{\uparrow,\downarrow}$, polarization $P$, magnetic susceptibility $\chi$
and specific heat $c_V$ which can be derived from the grandcanonical potential
(\ref{Ggas})
\[
n_{\uparrow,\downarrow}=-\frac{1}{2}\left(\frac{\6\phi}{\6\mu}\pm\frac{\6\phi}{\6 H}\right)\, ,\
P=\frac{n_\uparrow-n_\downarrow}{n_\uparrow+n_\downarrow}\, ,\
\chi=-\left[\left(\frac{\6^2 \phi}{\6 H^2}\right)_\mu+\left(\frac{\6 n}{\6 H}\right)^2_\mu\left(\frac{\6 n}{\6 \mu}\right)^{-1}_H\right]\, ,\
c_V=-T\left[\frac{\6^2\phi}{\6 T^2}+\left(\frac{\6 n}{\6 T}\right)
^2_\mu\left(\frac{\6 n}{\6\mu}\right)_T^{-1}\right]\, .
\]

At zero temperature and zero magnetic field the groundstate of the Gaudin-Yang
model is balanced (number of particles with spin up is equal with the number
of particles with spin down) and the excitation spectrum is gapless which
means that switching a magnetic field the system will become partially
polarized.
The monotonously decreasing polarization of the system as a function of the
reduced temperature  in the presence of a constant magnetic field is presented
in the upper panels of Fig.~(\ref{TData}).
We can also see that for a given $H$ and temperature the polarization is an
increasing function of the interaction strength. For large values of the dimensionless coupling strength the system
becomes paramagnetic as it can be seen from Eq.~(\ref{impf}) which describes
free fermions at chemical potential $\mu'=\mu +\ln[2\cosh(\beta H)]$.

The magnetic susceptibility is finite everywhere except at $\gamma=\infty,$ $H=0$ and $T=0$.
In this case we can see from Eq.~(\ref{impf})) that the magnetization is $M=n \tanh\left(H/T\right)$
and the susceptibility $\chi=n [1-\tanh^2 \left(H/T\right)]/T$ diverges like
$1/T$ at vanishing magnetic field. At low-temperatures and small magnetic fields $\chi$  presents a complex behavior
as a function of $\gamma$. For $H=1$ the maximum susceptibility is obtained for $\g=\infty$ but
already for $H=2.5$ the susceptibilities for $\g=\{0,1,3\}$ present more
pronounced maxima at very low-temperatures while the maximum for
$\chi(\g=\infty)$ moves at a higher temperature. For strong magnetic fields
the susceptibility is zero at low-temperatures,  presents a global maximum  at a
reduced temperature  close to the value of the magnetic field  and the dependence
on the coupling strength becomes small.  The specific heat is linear at low-temperatures
and similar to the case of the susceptibility presents a maximum at a temperature
$\tau \sim H$. A similar behavior was
noticed by Klauser and Caux \cite{KC} for the 2CBG but in our case the maxima
are more pronounced for the same values of temperatures and magnetic field. At
high temperatures the specific heat per particle approaches $1/2$, the value
for the ideal gas. As a function of the coupling strength and for low values
of the magnetic field the specific heat reaches its maximum for
$\gamma=\infty$ but this is no longer true for strong magnetic fields where
the dependence on $\g$ becomes less pronounced.

\section{Tan contact and local correlation functions}\label{s5}

S. Tan \cite{Tan1} discovered that the momentum distribution of 3D fermions
with zero range spin-independent interactions exhibits a universal $C/k^4$
behavior at large momenta (a similar relation for the Lieb-Liniger model was
derived earlier by Olshanii and Dunjko \cite{OD}). The large momentum distribution
is the same for both particle  species and the constant $C$,  called contact, is
given by the probability that  two particles of opposite spin are found at the same
position in space.
The contact also appears in Tan's
adiabatic theorem \cite{Tan1} which determines the change of the energy with
respect to the interaction strength and a series of universal thermodynamic
identities called Tan relations \cite{Tan1} (see also
\cite{BP1,ZL,WTC,BZ,VZM,WC}).  The analogues of Tan relations in 1D were
derived by Barth and Zwerger \cite{BZ} using the Operator Product Expansion of
Wilson and Kadanoff \cite{Wils,Kad}. The contact is not only an important and
valuable theoretical concept but is also experimentally measurable
\cite{NNCS,SGDJ,KHLDMD,KHHDHV2,KHDHHV1,SDPJ,WMPCJ,HLFHV}.

In the case of the Gaudin-Yang model with the Hamiltonian (\ref{Hc}) the
contact is defined as \cite{BZ} (remember $\hbar=2m=1$ and $c=-2/a_{1D}$
 with
$a_{1D}$ the scattering length)
\be
C=\frac{4}{a^2_{1D}}\int dx\, \langle \Psi_\uparrow^\dagger(x)\Psi_\downarrow^\dagger(x)\Psi_\downarrow(x)\Psi_\uparrow(x)\rangle=c^2\int dx\, \langle \Psi_\uparrow^\dagger(x)\Psi_\downarrow^\dagger(x)\Psi_\downarrow(x)\Psi_\uparrow(x)\rangle\, ,
\ee
where by $\langle\ \rangle$ we denote the zero or finite temperature
expectation value.
In our definition we can see that $C$ is an extensive
variable and is closely related to the interaction energy $ I=2c\int dx\,
\langle\Psi_\uparrow^\dagger(x)\Psi_\downarrow^\dagger(x)\Psi_\downarrow(x)\Psi_\uparrow(x)\rangle$
and the local correlation function of opposite spins defined by
\be
g_{\uparrow\downarrow}^{(2)}(x)=4  \frac{\langle\Psi_\uparrow^\dagger(x)\Psi_\downarrow^\dagger(x)\Psi_\downarrow(x)\Psi_\uparrow(x)\rangle}{n(x)^2}=
4  \frac{\langle\Psi_\uparrow^\dagger(x)\Psi_\downarrow^\dagger(x)\Psi_\downarrow(x)\Psi_\uparrow(x)\rangle}{(n_\uparrow(x)+n_\downarrow(x))^2}\, .
\ee
A simple application of the Hellmann-Feynman theorem gives the 1D version of
Tan's adiabatic theorem \cite{BZ} $\frac{dE}{da_{1D}}=
\left\langle\frac{\6\mathcal{H}_{GY}}{\6a_{1D}}\right\rangle=C$ which
expresses the variation of the total energy $E$ with respect to the scattering
length.

The situation is considerably simplified in the homogeneous case $V(x)=0$. The local correlation function and density no
longer depend on $x$ and we can introduce the  contact and interaction energy per length
\be
\mathcal{C}=\frac{c^2}{4} n^2g_{\uparrow\downarrow}^{(2)}(0)\, ,\ \ \  \mathcal{I}=\frac{c}{2}n^2g_{\uparrow\downarrow}^{(2)}(0)\, .
\ee
Another application of the Hellmann-Feynman theorem allows us to obtain the
local correlation function from the derivative of the grandcanonical potential
per length with respect to the coupling strength
$g_{\uparrow\downarrow}^{(2)}(0)=\frac{2}{n^2}\frac{\6 \phi}{\6 c}$.  Noting
that $\phi=-\ln \mathcal{Z}/(\beta L)$ with $\mathcal{Z}=Tr[e^{-\beta
    (\mathcal{H}_{GY}-\mu_\uparrow N_\uparrow-\mu_\downarrow N_\downarrow)}]$
we have $\frac{\6 \phi}{\6 c}=Tr[\frac{\6 \mathcal{H}_{GY}}{\6 c}e^{-\beta
    (\mathcal{H}_{GY}-\mu_\uparrow N_\uparrow-\mu_\downarrow
    N_\downarrow)}]/(L\mathcal{Z})$
$=2\langle\Psi_\uparrow^\dagger(0)\Psi_\downarrow^\dagger(0)\Psi_\uparrow(0)\Psi_\downarrow(0)\rangle_T$
which proves the identity.  For the homogeneous system the pressure and total
energy per length $\mathcal{E}$ satisfy the following Tan relation \cite{BZ}
$p=2\mathcal{E}-2\mathcal{C}/c$. We have checked numerically this relation and
found perfect agreement providing an additional check for the validity of our
NLIEs. Other analytical or numerical computations of the contact in 1D systems
can be found in \cite{HJLPAD,BAD,RPLD,YB,ZWKGD,DK,BBF,VM}.

\begin{figure}[ht]
\includegraphics[width=1\linewidth]{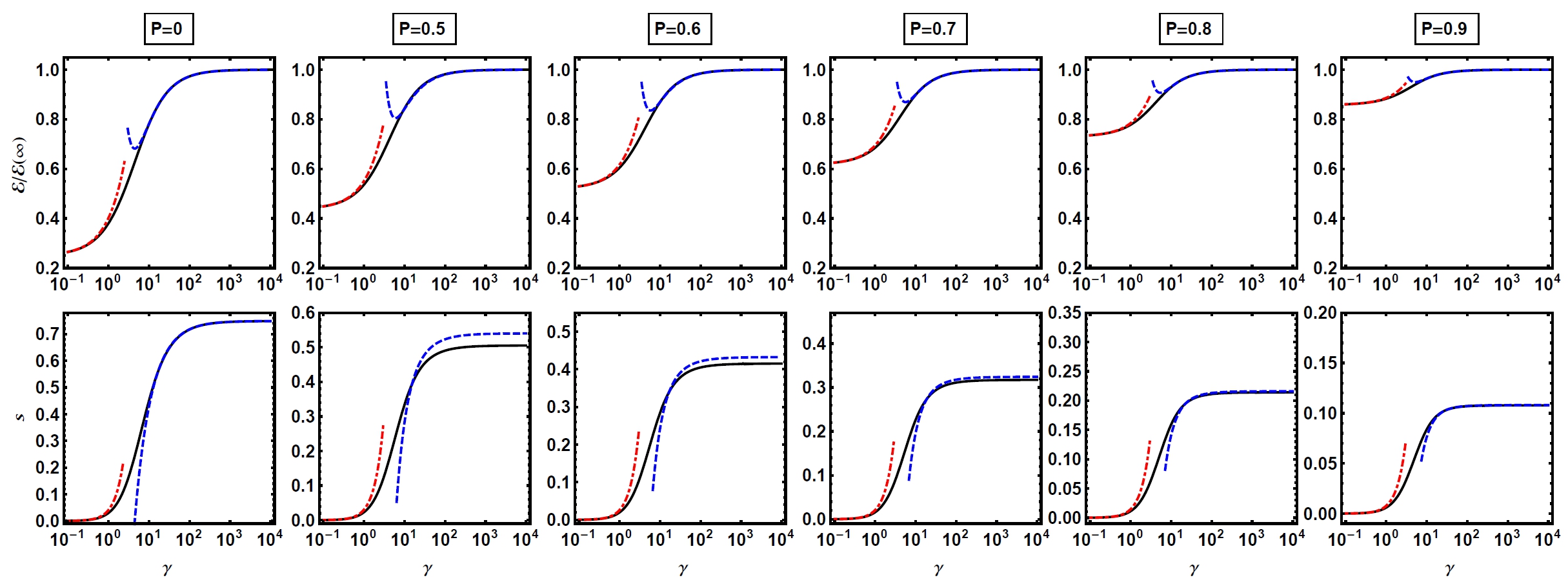}
\caption{(Color online) Plots of the zero temperature energy density
  $\mathcal{E}$ at fixed polarization (upper panels black continuous line) in
  units of $\mathcal{E}(\infty)=n^3 \pi^2/3$ and dimensionless contact
  $s=\mathcal{C}/k_F^4$ (lower panels black continuous line) with $k_F=\pi
  n/2$ as functions of the dimensionless coupling constant $\g.$ The dashed
  blue and dot-dashed red lines represent the asymptotic expansions in the strong
  and weak interaction limit given by Eqs.~(\ref{TGexp}), (\ref{Weakexp}) and
  (\ref{exp1}). The results presented in the top row can be seen as complementing  Fig.~1 of \cite{GM1}.}
\label{Tan0}
\end{figure}

\subsection{Results at T=0}

At zero temperature the thermodynamic properties of the system can be
extracted from the following system of Fredholm type integral equations  \cite{Y1}
\footnote{ An analytical derivation of Eqs.~(\ref{GS}) from the low-temperature limit of our NLIEs, Eqs.~(\ref{NLIEgas}), is still lacking.
The main difficulty lies in the fact that our (complex) auxiliary functions $a_{1,2}(k)$  are not easily related to the dressed energies of the TBA
formalism.}
\begin{subequations}\label{GS}
\begin{align}
\rho_c(k)&=\frac{1}{2\pi}+\int_{-\lambda_0}^{\lambda_0}b_1(k-\lambda)\rho_s(\lambda)\, d\lambda\, \\
\rho_s(\lambda)&=\int_{-k_0}^{k_0}b_1(\lambda-k)\rho_c(k)\, dk-\int_{-\lambda_0}^{\lambda_0}b_2(\lambda-\mu)\rho_s(\mu)\, d\mu
\end{align}
\end{subequations}
where $b_m(k)= m c/[2\pi[(mc)^2/4+k^2]].$ In Eqs.~(\ref{GS}) $k_0$ and
$\lambda_0$ are parameters which fix the values of the density of spin down
particles and energy per length via $n=\int_{-k_0}^{k_0} \rho_c(k)\, dk\, ,$
$n_\downarrow=\int_{-\lambda_0}^{\lambda_0} \rho_s(\lambda)\, d\lambda\, ,$
and $\mathcal{E}=\int_{-k_0}^{k_0} k^2\rho_c(k)\, dk\, .$ The balanced system
is characterized by $\lambda_0=\infty$ and the fully polarized system by
$\lambda_0=0$.  Even though in the general case an analytic solution of the
system of equations (\ref{GS})  is not known, in certain limits
approximate results can be derived. In the strong interaction limit ($\g\gg
1$) Guan and Ma \cite{GM1} obtained (the leading order term was derived in
\cite{Fuchs,BCS2})
\be\label{TGexp}
\mathcal{E}=\left\{\begin{array}{ll} \frac{n^3 \pi^2}{3}\left[1-\frac{4 \ln 2}{\g}+\frac{12(\ln 2)^2}{\g^2}-
                                         \frac{32(\ln 2)^3}{\g^3}+\frac{\pi^2\zeta(3)}{\g^3}\right]\, ,\ \ P=0\, , \\
                            \frac{n^3 \pi^2}{3} \left[1-\frac{4 (1-P)}{\g}+\frac{12(1-P)^2}{\g^2}
                             -\frac{32(1-P)^3}{\g^3}+\frac{16\pi^2(1-P)}{5\g^3}\right]\, ,\ \ P\ge 0.5\, ,
                 \end{array} \right.
\ee
with $\zeta(z)$ the Riemann function, and in the weak interaction limit $(\g\ll 1)$
\be\label{Weakexp}
\mathcal{E}=\frac{n^3 \pi^2(1-P)^3}{24}+  \frac{n^3 \pi^2(1+P)^3}{24}+\g\frac{n^3(1-P^2)}{2} \, .
\ee
It should be emphasized that Eqs.~(\ref{TGexp}) and (\ref{Weakexp}) are not exact. They represent numerical
approximations which are valid only in the Tonks-Girardeau  ($\g\gg 1$) and weak interaction ($\g\ll 1$)  regimes.
Using Tan's adiabatic theorem (see above) the local correlation function of
opposite spins, contact and interaction energy can be computed as
\be\label{exp1}
g_{\uparrow,\downarrow}^{(2)}(0)=\frac{2}{n^3}\frac{d\mathcal{E}}{d\g}\, ,
\ \ \mathcal{C}=\frac{n}{2} \g^2 \frac{d\mathcal{E}}{d\g}\,
,\ \ \ \mathcal{I}=\g \frac{d\mathcal{E}}{d\g}\, .  \ee In the following it
will be preferable to work with the dimensionless contact density defined as
$s=\mathcal{C}/k_F^4$ with $k_F=\pi n/2$.

In Fig.~\ref{Tan0} we present results for the zero temperature energy and
dimensionless contact as functions of the coupling constant obtained from the
numerical integration of Eqs.~(\ref{GS}) and the asymptotic expansions
(\ref{TGexp}) and (\ref{Weakexp}).  In the balanced case ($P=0$) the
asymptotic expansions for energy and contact represent a good approximation in
the weak and strong interacting regime but in the case of imbalanced systems
($P\ne 0$) the results derived from Eq.~(\ref{TGexp}) for the strongly
interacting regime become more accurate as the polarization approaches
one. The contact is a monotonously increasing function of $\gamma$ and in the
Tonks-Girardeau limit ($\g\rightarrow\infty$) approaches asymptotically a
finite value. As a function of the polarization $s$ reaches a maximum for
$P=0$ and is zero for the fully polarized system. For the balanced system in
the strongly interacting regime, using (\ref{TGexp}) and (\ref{exp1}) we
obtain
\be\label{Tanexp}
s(\gamma,P=0)=\frac{8}{3 \pi^2}\left[4 \ln 2-\frac{24(\ln 2)^2}{\g}+
                                         \frac{96(\ln 2)^3}{\g^2}-\frac{3 \pi^2\zeta(3)}{\g^2}\right]\, , \ \g\gg 1\, ,
\ee
with $s(\infty,P=0)=32 \ln 2/(3\pi^2)$ a result which was first
obtained in \cite{BZ}.

\begin{figure}[t]
\includegraphics[width=0.9\linewidth]{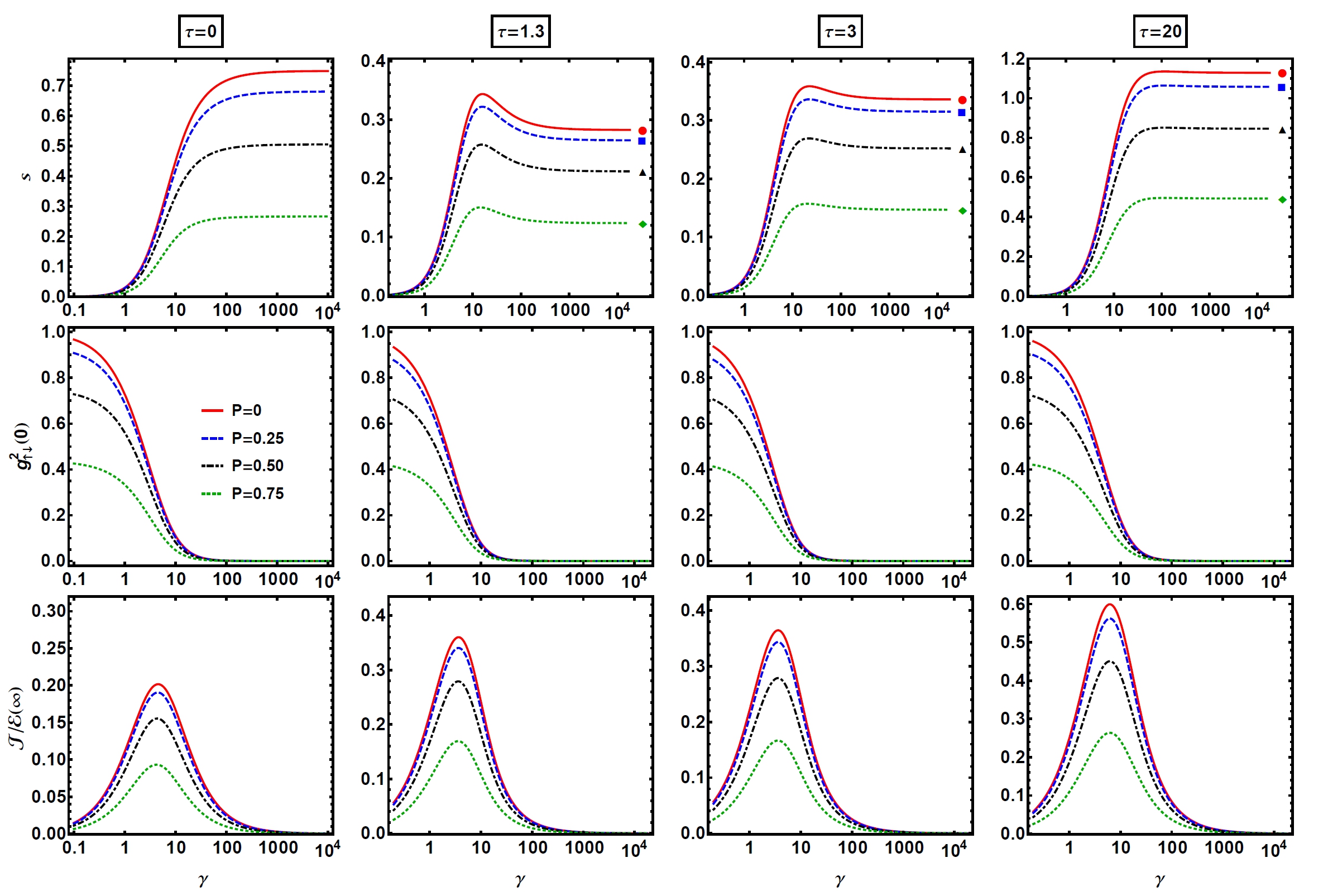}
\caption{(Color online) Plots of the dimensionless contact $s$ (upper panels)
  local correlation function of opposite spins
  $g_{\uparrow\downarrow}^{(2)}(0)$ (middle panels) and interaction energy
  $\mathcal{I}$ (lower panels) as functions of the dimensionless coupling
  constant $\g$ for $\tau=\{0, 1.3, 3, 20\}$ and fixed polarization $P=0$ (red
   line), $P=0.25$ (blue dashed line), $P=0.5$ (black dot-dashed line) and $P=0.75$ (green dotted
  line). The red disks, blue squares, black triangles and green diamonds
  represent the values of the dimensionless contact $s(\g=\infty)$ computed
  from the momentum distribution derived from the Fredholm determinants
  (\ref{Fredholm}) for $P=\{0, 0.25, 0.5, 0.75\}$.  The total density
    is fixed to $n=1/2$.}
\label{TanP}
\end{figure}

\subsection{Results at finite temperature}

As we have mentioned before at finite temperature analytical results are
restricted to the low-temperature strong-coupling regime \cite{LGSB} and
therefore we will have to rely on numerical solutions of the NLIEs
(\ref{NLIEgas}). In Fig.~\ref{TanP} we present the dependence on the
interaction strength of the dimensionless contact, local correlation function
and interaction energy for fixed polarization and several values of the
reduced temperature.

An interesting feature revealed by our data is that the contact at low and
intermediate temperatures (see upper panels of Fig.~\ref{TanP}) presents a
local maximum (for any value of the polarization) which does not appear at
$\tau=0$ and gets suppressed at high temperatures. In the Tonks-Girardeau
regime the contact approaches a finite value for all values of
$\tau$. Remembering that the contact governs the tail of the momentum
distribution ($\tilde n(k)\sim \mathcal{C}/k^4$ for $k\rightarrow\infty$) this
means that at $\tau=0$ and fixed polarization (or magnetic field) the width of
the momentum distribution increases monotonically with $\gamma$ reaching its
maximum at $\gamma=\infty.$ In contrast, at low and intermediate temperatures
the width of the momentum distribution reaches a maximum at a finite value of
the coupling strength and then becomes narrower as $\gamma\rightarrow\infty.$
This reconstruction of the momentum distribution as a function of the strength
of the interaction should be in principle experimentally accessible.

As shown in the middle panels of Fig.~\ref{TanP} the local correlation
function which quantifies the probability that two particles of opposite spin
occupy the same point in space is a monotonically decreasing function of
$\gamma$ with a maximum in the noninteracting limit
($g_{\uparrow\downarrow}^{(2)}(0)=1$ for $P=\g=0$ and all $\tau$). In the
strongly interacting limit $g_{\uparrow\downarrow}^{(2)}(0)$ drops to zero
like $1/\g^2$. For a fixed value of $\gamma$ as a function of the polarization
the local correlator is maximal  in the balanced system and vanishes in the fully
polarized case. The interaction energy (bottom row Fig.~\ref{TanP}) exhibits a
local maximum at zero and finite temperature and vanishes for
$\gamma\rightarrow \infty$.  This is due to the fact that in the strong
interaction limit the short-range repulsive interaction acts as an effective
Pauli exclusion principle also between fermions of opposite spins
\cite{BZ}. As expected, the interaction energy is largest for the $P=0$ case
and it is zero for the fully polarized system.

An alternative way of deriving the contact requires the analytical or
numerical computation of the momentum distribution which is the
Fourier transform of the static Green's function i.e., $\tilde n(k)=\inti
e^{-i k x} \langle \Psi_\uparrow^\dagger
(x)\Psi_\uparrow(0)\rangle_{\mu,H,T}\, dx.$ In the general case this is an
almost hopeless task due to the fact that our knowledge of the correlators is
still rather limited \cite{BL,Ber,ISLL1,IP1,GIKP,GIK,GKa}. However, in the
impenetrable case we can use the Fredholm determinant representations derived
by Izergin and Pronko \cite{IP1} which have the advantage of being extremely
easy to implement numerically using the method presented in \cite{Bor}. In our
case the Green's functions have the following representation \cite{IP1}
\be\label{Fredholm}
\langle \Psi_\uparrow^\dagger (x)\Psi_\uparrow(0)\rangle_{\mu,H,T}=\mbox{\tt det} \left(\hat{\textsf{I}}+\frac{e^{2H/T}}{2}\  \hat{\textsf{V}} +\hat\R \right)
-\mbox{\tt det} \left(\hat{\textsf{I}}+\frac{e^{2H/T}}{2} \ \hat{\textsf{V}}\right)\, ,\
\ee
where the integral operators $\hat{\textsf{V}}$ and $\hat{\textsf{R}}$ act on
an arbitrary function like $(\hat{\textsf{V}} f)(\lambda)=\inti \textsf{V}
(\lambda,\mu) f(\mu)\, d\mu$ and $(\hat{\textsf{R}} f) (\lambda)=\inti
\textsf{R} (\lambda,\mu) f(\mu)\, d\mu$ with kernels
\be
\textsf{V}(\lambda,\mu)=-\, \sqrt{\theta(\lambda)}\, \frac{2\sin\left(|x|\frac{\lambda-\mu}{2}\right)}{\pi(\lambda-\mu)}\, \sqrt{\theta(\mu)}\, ,\ \
\textsf{R}(\lambda,\mu)=-\, \sqrt{\theta(\lambda)}\, \frac{e^{- i x \frac{\lambda+\mu}{2}}}{2\pi}\, \sqrt{\theta(\mu)}\, ,
\ee
and $\theta(\lambda)=e^{-H/T}/(2\cosh(H/T)+e^{(\lambda^2-\mu)/T})$ is the
Fermi weight. In addition we have $\langle \Psi_\downarrow^\dagger
(x)\Psi_\downarrow(0)\rangle_{\mu,H,T}=\langle \Psi_\uparrow^\dagger
(x)\Psi_\uparrow(0)\rangle_{\mu,-H,T}\, .$ The results for the contact
obtained from the large $k$ analysis of the Fourier transform of
(\ref{Fredholm}) are presented in the $\tau=\{1.3,3,20\}$ panels of
Fig.~\ref{TanP} as red disks ($P=0$), blue squares ($P=0.25$), black triangles
($P=0.5$) and green diamonds ($P=0.75$) and they are in perfect agreement with
the results derived from the NLIEs.
\begin{figure}[t]
\includegraphics[width=1\linewidth]{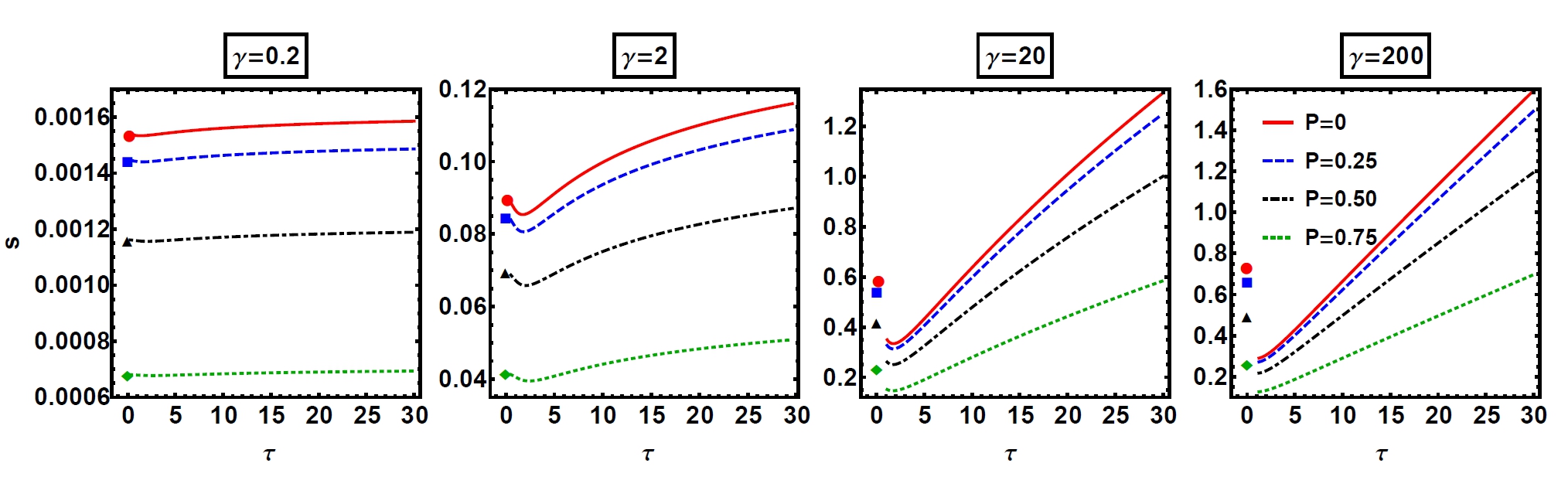}
\caption{(Color online) Reduced temperature dependence of the dimensionless
  contact $s$ for $\g=\{0.2, 2, 20, 200\}$ and fixed polarization $P=0$ (red
  line), $P=0.25$ (blue dashed line), $P=0.5$ (black dot-dashed line) and $P=0.75$ (green dotted
  line). The red disks, blue squares, black triangles and green diamonds
  represent the values of the dimensionless contact $s(\tau=0)$ computed from
  the integral equations (\ref{GS}) for $P=\{0, 0.25, 0.5, 0.75\}$.  }
\label{TanT}
\end{figure}

The temperature dependence of the contact shown in Fig.~\ref{TanT} reveals
another interesting phenomenon (in Fig.~\ref{TanT} the values of the contact
at $\tau=0$ were computed from the integral equations (\ref{GS}) and for
$\tau>0$ from the NLIEs (\ref{NLIEgas})).  For weak interactions the
dependence of $s$ on temperature is very small, but as we increase the
coupling strength, the contact develops a local minimum which becomes very
pronounced in the Tonks-Girardeau limit. The nonmonotonic behavior is present
for all polarizations and the temperature interval in which it
manifests itself shrinks as the value of $\g$ increases. Outside of this
temperature interval and for strong interactions the contact presents a linear
dependence on the reduced temperature for all values of $P.$ The nonmonotonic
behavior of the contact implies that the shape of the momentum distribution of
the strongly repulsive Gaudin-Yang model suffers an abrupt change in a very
small interval of temperature. This reconstruction of the distribution
 is also counterintuitive because as the temperature increases the
distribution becomes narrower and not wider as one would expect from a system
of weakly interacting or free particles. However, one should not forget that
in this case we are dealing with a strongly interacting system where this type
of picture  is not correct.

This crossover of the momentum distribution was first discovered by Cheianov,
Smith and Zvonarev in \cite{CSZ} for the balanced system and signals the
transition from the Tomonaga-Luttinger liquid phase to the incoherent spin
Luttinger liquid regime. Our results show that a similar crossover is present
also in the imbalanced case and it might be possible to be detected even for
moderate values of the coupling strength. Compared with the single-component
case in the two-component system there are two temperature scales $T_F=\pi^2
n^2$ and $T_0=T_F/\g$ (remember $\hbar=2m=k_B=1$) which define two different
quantum regimes: $T<T_0$ characterized by the Tomonaga-Luttinger liquid
theory and $T_0\ll T\ll T_F$ in which the incoherent spin Luttinger theory is
applicable. The two temperature scales are well separated in the strong
coupling limit and for $T_0\ll T\ll T_F$ the charge degrees of freedom are
effectively frozen while the spin degrees of freedom are strongly
``disordered".  It is important to note that in the computation of the
correlation functions the limits $c \rightarrow 0$ and $T\rightarrow 0$ do not
commute \cite{ISLL1}. The determinant representation (\ref{Fredholm}) has been
derived by taking the $c\rightarrow \infty$ limit in the wavefunctions
followed by the summation of form factors at finite temperature. If we take
the limit $T\rightarrow 0$ in (\ref{Fredholm}) and compute the dimensionless
contact from the tail of the momentum distribution we obtain
$s(\g=\infty,P=0)_{T>T_0}=0.270$ which is considerably smaller than
$s(\g=\infty,P=0)_{T<T_0}=32 \ln 2/(3\pi^2)=0.749...$ (see Eq.~(\ref{Tanexp}))
obtained from the integral equations at zero temperature (Eqs.~\ref{GS}).  The
momentum crossover for the balanced impenetrable system is presented in Fig.~1
of \cite{CSZ}.

\section{Density profiles at finite temperature}\label{s6}

In most experiments the system is subjected to a harmonic potential
$V(x)=m\omega^2 x^2/2$, a situation in which the Hamiltonian
(\ref{Hc}) is no longer exactly solvable. However, for slowly varying
potentials we can apply the Local Density Approximation (LDA) and the $V(x)=0$
solution to obtain reasonably accurate data for the trapped system. Under LDA
each point in the trap can be seen as a locally homogeneous system with
\be\label{i20} \mu(x)=\mu(0)-V(x)=\mu(0)-m\omega^2 x^2/2\, ,\ \ \ H(x)=H(0)\,
, \ee where $\mu(0)$ and $H(0)$ are the chemical potential and magnetic field
in the center of the trap. An immediate consequence of (\ref{i20}) is that the
density along the trap is monotonously decreasing and at zero temperature
vanishes at a distance $R_{TF}$ from the center of the trap where $R_{TF}$ is
called the Thomas-Fermi radius and is determined by
$\mu(0)-m\omega^2R_{TF}^2/2=0$. An inhomogeneous system can be characterized
by three parameters: the dimensionless coupling strength $\gamma(0)=c/n(0)$,
the polarization $P(0)=(n_\uparrow(0)-n_\downarrow(0))/n(0)$ and the reduced
temperature $\tau(0)=T/n^2(0)$ all evaluated at the center of the trap.
\begin{figure}[h]
\includegraphics[width=0.9\linewidth]{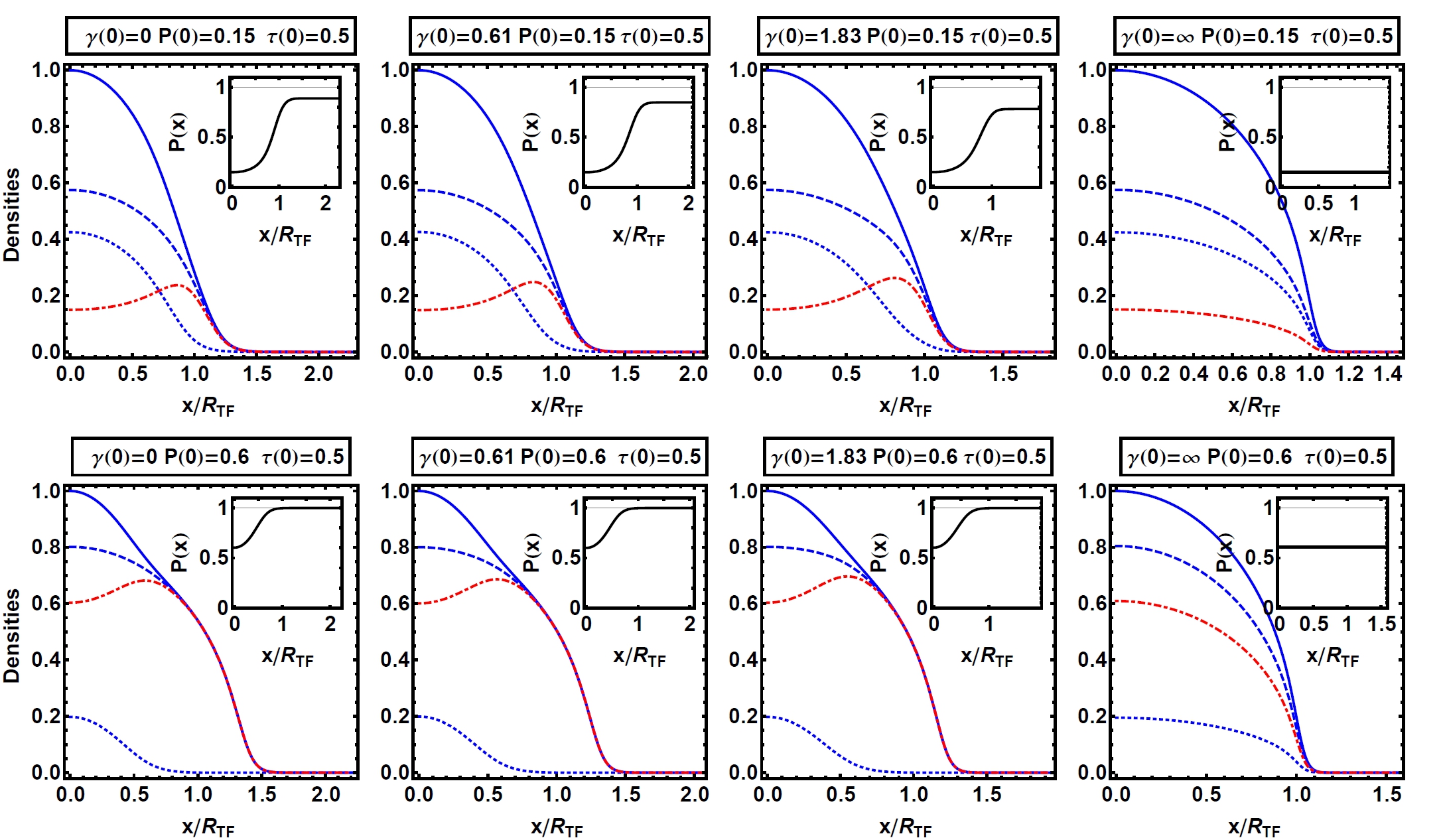}
\caption{(Color online)
 Plots of the normalized total density $n(x)/n(0)$
  (blue continuous line), normalized density of up spins $n_\uparrow(x)/ n(0)$ (blue dashed line),
  normalized density of down spins $n_\downarrow(x)/n(0)$ (blue dotted line) and their difference
  $(n_\uparrow(x)-n_\downarrow(x))/n(0)$
  (red dot-dashed line) for an inhomogeneous Gaudin-Yang model with
  $\tau(0)=0.5$ and polarizations $P(0)=0.15$ (upper panels) and $P(0)= 0.6$
  (lower panels). The insets depict the variation of the polarization
  $P(x)=(n_\uparrow(x)-n_\downarrow(x)) /(n_\uparrow(x)+n_\downarrow(x))$ in the
  trap. For all values of the coupling strength $\g(0)=\{0, 0.61, 1.83, \infty
  \}$ the total density in the center of the trap is $n(0)=\sqrt{24}$.  }
\label{DP}
\end{figure}
The density profiles at $T=0$ were computed in \cite{ColTat} and it was
noticed that for all $P(0)\in(0,1)$ they present a two shell structure: an
imbalanced mixture of spin up and down fermions in the center and fully
polarized wings. At finite temperature the situation becomes more complex as
it can be seen in Fig.~\ref{DP} where we present the density profiles at
$\tau(0)=0.5$ for different values of the coupling strength and two
polarizations. For $P(0)=0.15$ and $\tau(0)=0.5$ the entire system is in the
mixed phase for all values of $\gamma(0)$ (the density $n(0)=\sqrt{24}$ is the
same in all cases presented) even though the polarization along the trap
increases away from the center. The double shell structure appears only for
values of the center polarization above a critical value which is a monotonic
increasing function of $\gamma(0)$ and temperature (see the lower panels of
Fig.~\ref{DP}). At high temperatures the entire system will be found only in
the mixed phase irrespective of the value of $\gamma(0)$ and $P(0)$.

\section{The Gaudin-Yang model as the continuum limit of the Perk-Schultz spin chain with $(-++)$ grading}\label{s7}

The derivation of Eqs.~(\ref{Ggas}) and (\ref{NLIEgas}) is performed in three
steps. First, we identify a lattice embedding of the Gaudin-Yang model for
which the quantum transfer matrix method \cite{MS,K1} can be employed. The
need for this lattice model stems from the fact that the quantum transfer
matrix method, which has the advantage of producing a finite number of NLIEs
for the thermodynamics of the system, can be defined only for discrete
systems.  In the second step we calculate the free energy of this lattice
model from the largest eigenvalue of the QTM. Finally, the thermodynamics of
the continuum model is obtained by performing the scaling limit in the result
for the lattice model.

As in the case of the two-component Bose gas \cite{KP1,KP2} the appropriate
lattice model is the $q=3$ critical Perk-Schultz spin chain \cite{dVL,L}
\begin{align}\label{Hs}
\cal{H}_{PS}&=J\varepsilon_1\sum_{j=1}^L\left(\cos\g\sum_{a=1}^3\varepsilon_a\,  e_{aa}^{(j)}e_{aa}^{(j+1)}+
\sum_{\substack{a,b=1\\ a \ne b}}^3 e_{ab}^{(j)}e_{ba}^{(j+1)}
+i\sin\g\sum_{\substack{a,b=1\\ a\ne b}}^3 \mbox{sign}(a-b)e_{aa}^{(j)}e_{bb}^{(j+1)}\right)-\sum_{j=1}^L\sum_{a=1}^3 h_a e_{aa}^{(j)}\, ,
\end{align}
where $J>0$ is the coupling strength, $L$ is the number of lattice sites,
$\g\in[0,\pi]$ is the anisotropy, $h_1,h_2,h_3$ are chemical potentials and
$\varepsilon_1,\varepsilon_2,\varepsilon_3,\, (\varepsilon_i\in\{\pm 1\})$ are
the grading parameters.  Also $e^{(j)}_{ab}=\mathbb{I}_3^{\otimes j-1}\otimes
e_{ab}\otimes \mathbb{I}_3^{\otimes L-j}\, ,$ with $e_{ab}$ and $\mathbb{I}_3$
the canonical basis and the unit matrix in the space of $3$-by-$3$
matrices. If in the case of the 2CBG we considered the grading
$(\varepsilon_1,\varepsilon_2,\varepsilon_3)=(- - -),$ for the Gaudin-Yang
model the relevant grading is $(\varepsilon_1,\varepsilon_2,\varepsilon_3)=(-
+ +)$. The energy spectrum of the spin chain is (see Appendix \ref{a1})
\be\label{es}
E_{PS}=\sum_{j=1}^M e_0(v^{(1)}_j)+M_1(h_2-h_3)+E_0\, ,\ \ \
e_0(v)=J\frac{\sin^2\g}{\sin(v-\g)\sin v}+h_1-h_2\, ,\ \ \ E_0=JL\cos\g-h_1L\, ,
\ee
with the Bethe ansatz equations
\begin{subequations}\label{Bs}
\begin{align}
& \left(\frac{\sin(\g-v_s^{(1)})}{\sin v_s^{(1)}}\right)^L=(-1)^{M-1}
\prod_{p=1}^{M_1}\frac{\sin(v_s^{(1)}-v_p^{(2)}-\g)}{\sin(v_s^{(1)}-v_p^{(2)})}\, ,\ \ \   s=1,\cdots, M\, ,\\
&\prod_{j=1}^M\frac{\sin(v_l^{(2)}-v_j^{(1)}+\g)}{\sin(v_l^{(2)}-v_j^{(1)})}=
\prod_{\substack{p=1\\p\ne l}}^{M_1}\frac{\sin(v_l^{(2)}-v_p^{(2)}+\g)}{\sin(v_l^{(2)}-v_p^{(2)}-\g)}\, ,\ \ \  l=1,\cdots, M_1\, .
\end{align}
\end{subequations}

In order to prove that the Perk-Schultz spin chain with the $(-++)$ grading is
the lattice embedding of the 2CFG we are going to show that the spectrum and
Bethe equations of the continuum model (\ref{ef}), (\ref{BEc}) can be obtained
in a scaling limit from the lattice analogues (\ref{es}), (\ref{Bs}). We start
with the Bethe ansatz equations.  Performing the transformation
$v_s^{(1)}\rightarrow i\delta k_s^{(1)}/\epsilon +\g/2\, ,$ and
$v_l^{(2)}\rightarrow i\delta k_l^{(1)}/\epsilon +\pi/2$ with $\delta$ the
lattice constant and $\epsilon\rightarrow 0$ a small parameter,
Eqs.~(\ref{Bs}) take the form
\begin{align*}
&\left((-)\frac{\sinh(\delta k_s^{(1)}/\epsilon-i\g/2)}{\sinh(\delta k_s^{(1)}/\epsilon+i\g/2)}\right)^L=(-1)^{M-1}
\prod_{p=1}^{M_1}\frac{\cosh(\delta k_s^{(1)}/\epsilon-\delta k_p^{(2)}/\epsilon-i\g/2)}{\cosh(\delta k_s^{(1)}/\epsilon-\delta k_p^{(2)}/\epsilon+i\g/2)}\, ,\\
&\prod_{j=1}^{M}\frac{\cosh(\delta k_l^{(2)}/\epsilon-\delta k_j^{(1)}/\epsilon-i\g/2)}{\cosh(\delta k_l^{(2)}/\epsilon-\delta k_j^{(1)}/\epsilon+i\g/2)}
=\prod_{\substack{p=1\\p\ne l}}^{M_1}\frac{\sinh(\delta k_l^{(2)}/\epsilon-\delta k_p^{(2)}/\epsilon -i\g)}
{\sinh(\delta k_l^{(2)}/\epsilon-\delta k_p^{(2)}/\epsilon +i\g)}\, .
\end{align*}
Considering $\g=\pi-\epsilon$ we obtain
\begin{subequations}\label{i1}
\begin{align}
&\left(\frac{\cosh(\delta k_s^{(1)}/\epsilon+i\epsilon/2)}{\cosh(\delta k_s^{(1)}/\epsilon-i\epsilon/2)}\right)^L=(-1)^{M-1}
\prod_{p=1}^{M_1}(-)\frac{\sinh(\delta k_s^{(1)}/\epsilon-\delta k_p^{(2)}/\epsilon+i\epsilon/2)}{\sinh(\delta k_s^{(1)}/\epsilon-\delta k_p^{(2)}/\epsilon-i\epsilon/2)}\, ,\\
&\prod_{j=1}^{M}(-)\frac{\sinh(\delta k_l^{(2)}/\epsilon-\delta k_j^{(1)}/\epsilon+i\epsilon/2)}{\sinh(\delta k_l^{(2)}/\epsilon-\delta k_j^{(1)}/\epsilon-i\epsilon/2)}
=\prod_{\substack{p=1\\p\ne l}}^{M_1}\frac{\sinh(\delta k_l^{(2)}/\epsilon-\delta k_p^{(2)}/\epsilon +i\epsilon)}
{\sinh(\delta k_l^{(2)}/\epsilon-\delta k_p^{(2)}/\epsilon -i\epsilon)}\, .
\end{align}
\end{subequations}
If we take the limits $L\rightarrow\infty, \delta\rightarrow 0$, $(L\sim
\mathcal{O}(1/\epsilon^2)\, , \delta\sim \mathcal{O}(\epsilon^2))$ such that
$L\delta=L_F$, $c=\epsilon^2/\delta$, identifying $M_1$ with $M_\downarrow$
and using
\[
\frac{\cosh(\delta k_s^{(1)}/\epsilon+i\epsilon/2)}{\cosh(\delta k_s^{(1)}/\epsilon-i\epsilon/2)}\sim
\frac{1+ik_s^{(1)}\delta/2}{1-ik_s^{(1)}\delta/2}
\]
we see that for $M$ even and $M_1$ odd Eqs.~(\ref{i1}) transform in the Bethe equations of the Gaudin-Yang model (\ref{BEc}).

Under this set of transformations the energy spectrum becomes
\be\label{i2}
\beta(E_{PS}-E_0)=\sum_{j=1}^M\beta\left[J\delta^2(k_j^{(1)})^2-J\epsilon^2-J\epsilon^4/4\right]+
\beta(h_1-h_2)M+\beta(h_2-h_3)M_1+\mathcal{O}(\epsilon^6)\, .
\ee
If we denote by $\bar\beta$ the inverse temperature of the continuum model and
consider $\beta=\bar\beta/\delta^2$, $J=1,$ $h_1,h_2,h_3\rightarrow 0$
$(h_1\sim \mathcal{O}(\epsilon^2)\, , h_2,h_3 \sim \mathcal{O}(\epsilon^4)) $
such that $J\epsilon^2/ \delta^2-h_1/\delta^2$ is finite we have
\be\label{rel1}
e^{\beta E_0}Z(h_1,h_2,h_3,\beta)\rightarrow \mathcal{Z}(\mu,H,\bar\beta)\, .
\ee
where $Z(h_1,h_2,h_3,\beta)$ is the canonical partition function of the spin
chain and $\mathcal{Z}(\mu,H,\bar\beta)$ is the grandcanonical partition
function of the Gaudin-Yang model.
Therefore, we have showed that by performing the scaling limit presented above
(the spectral parameter $v\rightarrow i \delta k/\epsilon$) the thermodynamics
of the 2CFG can be derived from the similar result for the low-T critical
Perk-Schultz spin chain with $(-++)$ grading. It should be noted that the
scaling limit presented here is the same as the one employed in the 2CBG case
(see Table I of \cite{KP2}) the only difference being the grading.

\section{Free energy of the $(-++)$ Perk-Schultz spin chain }\label{s8}

The importance of the quantum transfer matrix \cite{MS,Koma,SI1,K1,K2} in the
study of integrable lattice models resides in the fact that not only  the
free energy of the model is related to the largest eigenvalue of the QTM but
also various correlation lengths can be derived from the spectrum of the
next-largest eigenvalues.  The definition of the QTM for the Perk-Schultz
spin chain in the algebraic Bethe ansatz framework can be found in
\cite{KP2}. The interested reader can find pedagogical introductions in the
subject in \cite{K3,GKS1,GSuz}.

The largest eigenvalue of the quantum transfer matrix (see Appendix \ref{a1}),
which will be denoted by $\Lambda_0(v)$, is found in the $(N/2,N/2)$-sector of
the spectrum which means that $m=n=N/2$ in Eqs.~(\ref{eigenvalueGM}) and
(\ref{BAEGM}).  $\Lambda_0(v)$ is related to the free energy per lattice site
of the Perk-Schultz spin chain via the relation $f(\beta,h_1,h_2,h_3)=-(\ln
\Lambda_0(0))/\beta$. In the following it will be useful to introduce the
notations $\phi_{\pm}(v)=\left(\sinh(v\pm iu)/\sin \g\right)^{N/2}$ with
$u=J\sin\g \beta/N$ where $N$ is the Trotter number and
$q_0(v)=\phi_-(v),\ q_j(v)=\prod_{r=1}^{N/2}\sinh(v-v_{r}^{(j)}), \ j=1,2,
\ q_3(v)=\phi_+(v).$ Changing the spectral parameter $v\rightarrow iv$ and
considering $N\in 4\mathbb{N}$, the expression for the largest eigenvalue of
the QTM (\ref{eigenvalueGM}) can be written as (see the last remark of
Appendix \ref{a1})
\be\label{i3}
\Lambda_0(v)=\lam_1(v)+\lam_2(v)+\lam_3(v)\, ,\ \ \ \lam_j(v)=\phi_-(v)\phi_+(v)\frac{q_{j-1}(v-i\tilde\varepsilon_j\g)}{q_{j-1}(v)}
\frac{q_{j}(v+i\tilde\varepsilon_j\g)}{q_{j}(v)}e^{\beta \tilde h_j}\, ,
\ee
with $(\tilde\varepsilon_1,\tilde\varepsilon_2,\tilde\varepsilon_3)=$ $(+-+)$
and $( \tilde h_1, \tilde h_2, \tilde h_3)=(h_3,h_1,h_2)$. In this notation
the BAEs (\ref{BAEGM}) take the form $
\ \lam_j(v^{(j)}_r)/\lam_{j+1}(v^{(j)}_r)=-1\, ,\ r=1,\cdots,N/2\, ,
(j=1,2). $

\subsection{Nonlinear integral equations for the auxiliary functions}

The derivation of the nonlinear integral equations and integral expression for
the largest eigenvalue of the QTM in the Gaudin-Yang model is very similar
with the one presented in \cite{KP2} for the two-component Bose gas. For this
reason, below we will not be as explicit as in the 2CBG case but we will
highlight the particular features introduced by the fermionic model.

\begin{figure}
\includegraphics[scale=0.15]{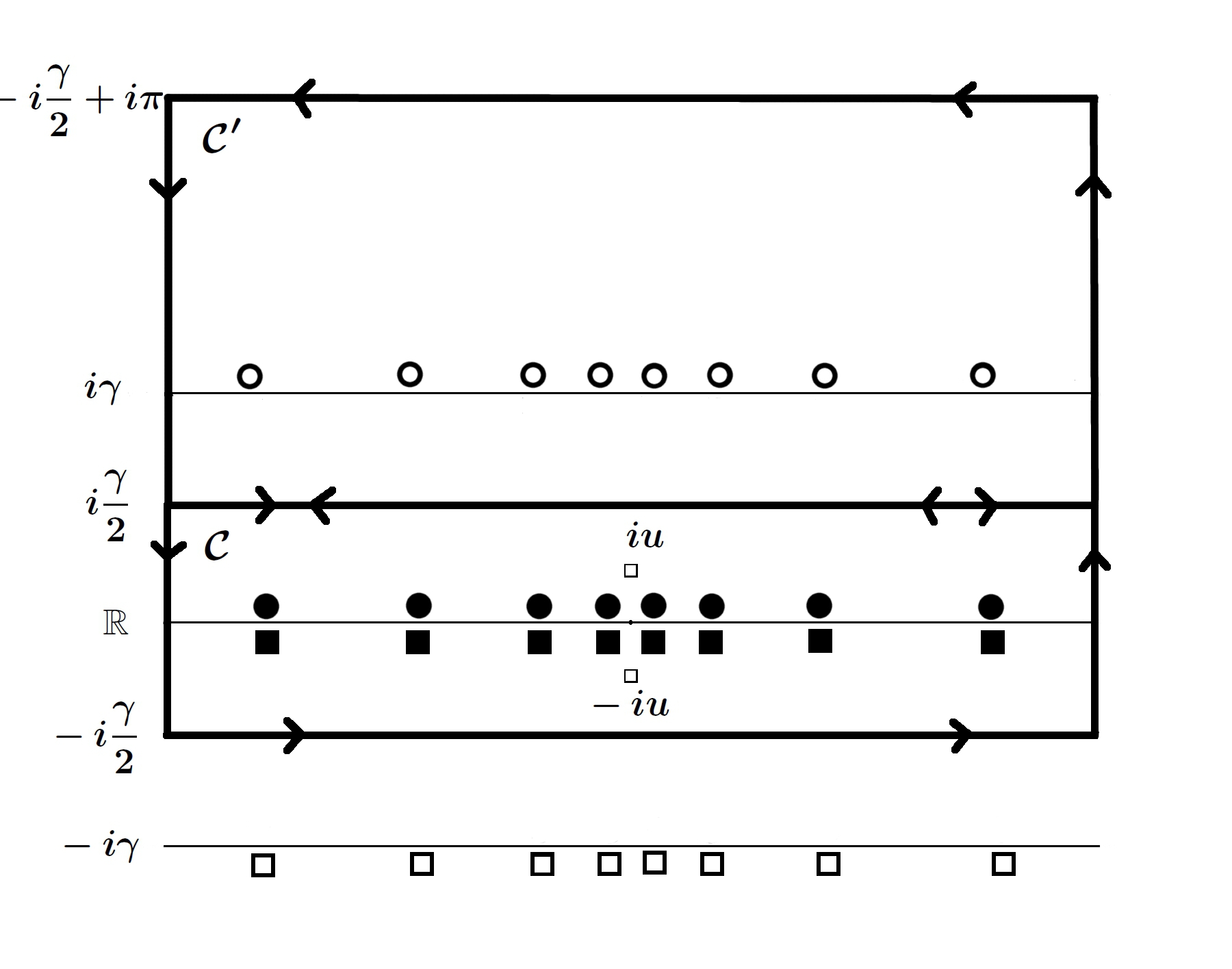}
\caption{Relevant contours and distribution of Bethe roots $\{v^{(1)}_i\}\,
  ,\{v^{(2)}_j\}$ ($\bullet,\blacksquare$) and holes $\{v'^{(1)}_i\}\,
  ,\{v'^{(2)}_j\}$ ($\circ,\square$) for the largest eigenvalue
  ($\g\in(0,\pi/2)$). The upper and lower edges of $\mathcal{C}$ are parallel
  with the real axis and intersect the imaginary axis at $\pm
  i(\g-\varepsilon) /2$ with $\varepsilon\rightarrow 0$.  }
\label{LE}
\end{figure}

We introduce two auxiliary functions periodic of period $i\pi$ defined by
\begin{subequations}\label{daux}
\begin{align}
\au(v)&=\frac{\lam_1(v)}{\lam_2(v)}=\frac{\phi_-(v-i\g)}{\phi_-(v)}\frac{q_2(v)}{q_2(v-i\g)}e^{\beta(h_3-h_1)}\, , \\
\ad(v)&=\frac{\lam_3(v)}{\lam_2(v)}=\frac{\phi_+(v+i\g)}{\phi_+(v)}\frac{q_1(v)}{q_1(v+i\g)}e^{\beta(h_2-h_1)}\, .
\end{align}
\end{subequations}
The equation $\mathfrak{a}_{1,2}(v)=-1$ has $N$ solutions and contains as a
subset the $N/2$ Bethe roots (see the remark after Eq.~(\ref{i3})) which will
be denoted by $\{v_{j}^{(1,2)}\}_{j=1}^{N/2}$. The additional $N/2$ solutions
are called holes and will be denoted by $\{v_{j}'^{(1,2)}\}_{j=1}^{N/2}.$ We
will first focus on the case when $\g\in(0,\pi/2)$ and we are going to assume
that the Bethe roots and holes for the largest eigenvalue of the QTM are
distributed in the complex plane as in Fig.\ref{LE}.
For $v$ outside the contour $\mathcal{C}$ we  introduce two additional functions $(j=1,2)$
\be\label{f}
f_j(v)\equiv\frac{1}{2\pi i}\int_{\mathcal{C}}\frac{d}{dv}\left(\ln\sinh(v-w)\right)
\ln(1+\mathfrak{a}_j(w))dw=\frac{1}{2\pi i}\int_{\mathcal{C}}\ln\sinh(v-w)
\frac{\mathfrak{a}_j'(w)}{1+\mathfrak{a}_j(w)}dw\, .
\ee
Let $g(v)$ be a function which is analytic inside and on $\mathcal{C}$.
Consider another function $\phi(v)$ which is also analytic inside the contour
with the exception of some poles.  Then (see pp. 129 of \cite{WW})
\be\label{thm} \frac{1}{2\pi
  i}\int_{\mathcal{C}}g(\la)\frac{\phi'(\la)}{\phi(\la)}d\la=
\sum_{i\in\mbox{zeros}}r_ig(a_i) -\sum_{i\in\mbox{poles}}s_ig(b_i)\, , \ee
where $r_i$ and $s_i$ are the multiplicities of the zeros $a_i$ and poles
$p_i$ of the function $\phi(v)$ inside the contour.  The application of
(\ref{thm}) with $g(v)=\ln\sinh(v-w)$ and $\phi(v)=1+\mathfrak{a}_j(v)$ gives
\begin{subequations}\label{i6}
\begin{align}
f_1(v)&=\ln q_1(v)-\ln \phi_-(v)-\frac{N}{2}\ln\sin\g\, ,\\
f_2(v)&=\ln q_2(v)-\ln \phi_+(v)-\frac{N}{2}\ln\sin\g\, .
\end{align}
\end{subequations}
Taking the logarithm of Eqs.~(\ref{daux}) and using (\ref{i6}) we find
\begin{subequations}\label{i7}
\begin{align}
\ln \au(v)=\beta(h_3-h_1)+\ln\left(\frac{\phi_-(v-i\g)}{\phi_+(v-i\g)}\frac{\phi_+(v)}{\phi_-(v)}\right)
+f_2(v)-f_2(v-i\g)\, ,\\
\ln \ad(v)=\beta(h_2-h_1)+\ln\left(\frac{\phi_+(v+i\g)}{\phi_-(v+i\g)}\frac{\phi_-(v)}{\phi_+(v)}\right)
+f_1(v)-f_1(v+i\g)\, .
\end{align}
\end{subequations}
The Trotter limit $N\rightarrow \infty$ can be taken in Eqs.~(\ref{i7})
($\lim_{N\rightarrow\infty} \ln\left(\phi_+(v)/\phi_-(v)\right)=i\beta
J\sin\g\coth v $) with the result
\begin{subequations}\label{nlie}
\begin{align}
\ln\au(v)&=\beta(h_3-h_1)-\beta \frac{J\sinh^2(i\g)}{\sinh v\sinh(v-i\g)}
-\int_{\mathcal{C}}K_2(v-w)\ln(1+\ad(w))\, dw,\label{nlie1}\\
\ln\ad(v)&=\beta(h_2-h_1)-\beta \frac{J\sinh^2(i\g)}{\sinh v\sinh(v+i\g)}+\int_{\mathcal{C}}K_1(v-w)\ln(1+\au(w))\, dw,
\label{nlie2}
\end{align}
\end{subequations}
where
\be
K_1(v)=\frac{1}{2\pi i}\frac{\sinh(i\g)}{\sinh(v+i\g)\sinh v}\, ,\ \
K_2(v)=\frac{1}{2\pi i}\frac{\sinh(i\g)}{\sinh(v-i\g)\sinh v}\, .
\ee
Eqs.~(\ref{nlie}) were derived assuming that $v$ and $v\pm i\g$ are located
outside the contour $\mathcal{C}$.  For $v$ inside the contour we should
add an additional term $\ln(1+\mathfrak{a}_2(v))$ on the r.h.s. of
(\ref{nlie1}) and $\ln(1+\mathfrak{a}_1(v))$ on the r.h.s. of (\ref{nlie2}).
The NLIEs (\ref{nlie}) remain valid also for $\g\in(\pi/2,\pi)$ but in this
case the upper and lower edges of the contour $\mathcal{C}$ are situated at
$\pm i(\pi-\g-\varepsilon)/2$ .

\subsection{Integral representation of the largest eigenvalue}\label{s82}

Due to the fact that the largest eigenvalue of the QTM is analytic in a strip
around the real axis we can calculate $\Lambda_0(v_0)$ for an arbitrary $v_0$
close to the real axis and then take the limit $v_0\rightarrow 0$ (the
expression for the free energy requires $\Lambda_0(0)$). Choosing $v_0=iu$ for
which $\lam_3(v_0)=0$ and using (\ref{id1}) we obtain
\be\label{temp}
\Lambda_0(v_0)= c\frac{\phi_+(v_0)}{q_2(v_0)}q_1(v_0+i\g)q_1^{(h)}(v_0)\, ,\ \  q_1^{(h)}(v)=\prod_{j=1}^{N/2}\sinh(v-{v'}_j^{(1)})\, ,
\ee
which highlights the need for an integral representation of $q_1^{(h)}(v_0)$
with $v_0$ close to the real axis ($\{{v'}_j^{(1)}\}_{j=1}^{N/2}$ are the
holes of $\mathfrak{a}_1(v)=-1$).  Let $v$ be a point close to the real axis
(inside the contour $\mathcal{C}$). As a result of (\ref{id2}) we have
($d(v-w)=\frac{d}{dv}\ln\sinh(v-w)$)
\[
\int_{\mathcal{C}}d(v-w)\frac{\mathfrak{a}_1'(w)}{1+\mathfrak{a}_1(w)}dw=
-\int_{\mathcal{C}'}d(v-w)\frac{\mathfrak{a}_1'(w)}{1+\mathfrak{a}_1(w)}dw\, ,
\]
where the r.h.s. can be evaluated with the help of (\ref{thm}). We find
(inside $\mathcal{C}$ the function $1+\au(w)$ has $N/2$ zeros at the holes
$\{v_{j}'^{(1)}\}_{j=1}^{N/2}$ and $N/2$ poles at $
\{v_{j}^{(2)}+i\g\}_{j=1}^{N/2}$)
\be\label{i8}
\int_{\mathcal{C}}d(v-w)\frac{\mathfrak{a}_1'(w)}{1+\mathfrak{a}_1(w)}dw=-\left(
\sum_{j=1}^{N/2}d(v-v_{j}'^{(1)})-\sum_{j=1}^{N/2}d(v-v_j^{(2)}-i\g)\right)\, .
\ee
If $v$ is close to the real axis then $v+i\g$ is outside the contour $\mathcal{C}$.
Therefore, with the help of (\ref{thm}) we can similarly compute
\be\label{i9}
\int_{\mathcal{C}}d(v-w+i\g)\frac{\mathfrak{a}_1'(w)}{1+\mathfrak{a}_1(w)}dw=\sum_{j=1}^N d(v-v_j^{(1)}+i\g)-\ln\phi_-\frac{N}{2}d(v-iu+i\g)\, .
\ee
Taking the difference of (\ref{i8}) and (\ref{i9}) and integrating by parts
w.r.t. $w$ and then integration w.r.t. $v$ we obtain
\be\label{i10}
\int_{\mathcal{C}}[d(v-w)-d(v-w+i\g)]\ln (1+\mathfrak{a}_1(w))dw=-\ln q_1^{(h)}(v)+\ln q_2(v-i\g)-
\ln q_1(v+i\g)+\ln\phi_-(v+i\g)+c\, ,
\ee
with $c$ a constant.
In an analogous fashion we can show that
\be\label{i11}
\int_{\mathcal{C}}[d(v-w)-d(v-w-i\g)]\ln (1+\mathfrak{a}_2(w))dw=-\ln q_2^{(h)}(v)+\ln q_1(v+i\g)-
\ln q_2(v-i\g)+\ln\phi_+(v-i\g)+c\, .
\ee
Taking the logarithm of (\ref{temp}) we obtain
\begin{align}
\ln \Lambda_0(v_0)&= \ln \phi_+(v_0) -\ln q_2(v_0)+\ln q_1(v_0+i\g)+\ln q_1^{(h)}(v_0)\, ,\\
&=\ln q_1^{(h)}(v_0)+\ln q_2^{(h)}(v_0)-\ln(1+\mathfrak{a}_2(v_0))+c\, ,
\end{align}
where in the last line we have used the identity (\ref{newi2}).
Using (\ref{i10}) and (\ref{i11}) we find
\begin{align*}
\ln \Lambda_0(v_0)=\ln(\phi_+(v_0-i\g)\phi_-(v_0+i\g))-\int_{\mathcal{C}}K_1(v_0-w)\ln(1+\au(w))\, dw+&\int_{\mathcal{C}}K_2(v_0-w)\ln(1+\au(w))\, dw\, \\
&-\ln(1+\mathfrak{a}_2(v_0))+c\, .
\end{align*}
The constant of integration can be obtained as in \cite{KP2} using the
behaviour of the involved functions at infinity with the result $c=\beta
h_1+\ln\left[\frac{1+e^{\beta (h_2-h_1)}+e^{\beta
      (h_3-h_1)}}{1+e^{\beta(h_3-h_1)}}\right]$.  Now, we can take the Trotter
limit followed by $v_0\rightarrow 0,$ $\lim_{\substack{v\rightarrow
    0,}{N\rightarrow \infty}}\ln(\phi_ +(v-i\g)\phi_-(v+i\g))=-J\beta\cos\g$,
obtaining the integral expression for the largest eigenvalue of the QTM
\be\label{le}
\ln\Lambda_0(0)=c-J\beta\cos\g-\int_{\mathcal{C}}K_2(w)\ln(1+\au(w))\, dw+\int_{\mathcal{C}}K_1(w)\ln(1+\ad(w))\, dw-\ln(1+\mathfrak{a}_2(0))\, .
\ee
For $\g\in(\pi/2,\pi)$ the same representation remains valid with the upper
and lower edges of the contour $\mathcal{C}$ situated at $\pm
i(\pi-\g-\varepsilon)/2$. We should mention that other thermodynamic
descriptions of the Perk-Schultz spin chain and related models can be found in
\cite{KWZ,PZinn,TT1,Tsub1}.

\section{Continuum limit}\label{s9}

In the continuum limit $\g=\pi-\epsilon.$ This means that in the NLIEs
(\ref{nlie}) and integral representation for the largest eigenvalue (\ref{le})
the upper and lower edge of the contour $\mathcal{C},$ denoted by
$\mathcal{C}_\pm$, are parallel with the real axis and intercept the imaginary
axis at $\pm i(\pi-\g-\varepsilon)/2$.  At low-temperatures and for
$J>0,\ h_3<h_1$ the driving term  on the r.h.s of
Eq.~(\ref{nlie1}) is large and negative on $\mathcal{C}_-$ which implies
that $\au(v)$ vanishes on the lower edge of the contour. In a similar way for
$h_2<h_1$ we can show that $\ad(v)$ is zero on $\mathcal{C}_+$. Therefore we
obtain
\begin{subequations}
\begin{align}
\ln\au(v)&=\beta(h_3-h_1)-\beta \frac{J\sinh^2(i\g)}{\sinh v\sinh(v-i\g)}
-\int_{\mathcal{C}_-}K_2(v-w)\ln(1+\ad(w))\, dw,\\
\ln\ad(v)&=\beta(h_2-h_1)-\beta \frac{J\sinh^2(i\g)}{\sinh v\sinh(v+i\g)}+\int_{\mathcal{C}_+}K_1(v-w)\ln(1+\au(w))\, dw\, ,
\end{align}
\end{subequations}
for the NLIEs and
\be
\ln\Lambda_0(0)=c-J\beta\cos\g+\int_{\mathcal{C}_+}K_2(w)\ln(1+\au(w))\, dw+\int_{\mathcal{C}_-}K_1(w)\ln(1+\ad(w))\, dw-\ln(1+\mathfrak{a}_2(0))\, ,
\ee
for the integral representation of the largest eigenvalue. Shifting the
argument $v$ and variable of integration to the line $+i\g/2$ ($-i\g/2$) for
the function $\au(v)$ ($\ad(v)$) we find
\begin{subequations}\label{i15}
\begin{align}
\ln\au(v+i\g/2)=\beta(h_3-h_1)&-\beta \frac{J\sinh^2(i\g)}{\sinh (v+i\g/2)\sinh(v-i\g/2)}
 -\int_{\mathbb{R}}K_2(v-w+i\g-i\varepsilon)\ln(1+\ad(w-i\g/2))\, dw,\\
\ln\ad(v-i\g/2)=\beta(h_2-h_1)&-\beta \frac{J\sinh^2(i\g)}{\sinh (v+i\g/2)\sinh(v-i\g/2)}-
\int_{\mathbb{R}}K_1(v-w-i\g+i\varepsilon)\ln(1+\au(w+i\g/2))\, dw\, ,\nonumber\\
\end{align}
\end{subequations}
and
\begin{align}\label{i16}
\ln\Lambda_0(0)=c-J\beta\cos\g+\int_{\mathbb{R}}K_2(w+i\g/2)\ln(1+\au(w+i\g/2))\, dw+&\int_{\mathbb{R}}K_1(w-i\g/2)\ln(1+\ad(w-i\g/2))\, dw\, \nonumber\\
&\ \ \ \ \ \ \ \ \ \ \ \ \ \ \ \ \ \ -\ln(1+\mathfrak{a}_2(0))\, .
\end{align}
In the scaling limit $v\rightarrow i\delta k/\epsilon\, , w\rightarrow i\delta
k'/\epsilon\, $ (the i factor is not needed because we have already changed
the spectral parameter to $iv$ in (\ref{i3})) $\g=\pi-\epsilon$ and \be
K_1(v-i\gamma)\rightarrow -\frac{\epsilon}{\delta}\frac{1}{2\pi
}\frac{c}{k(k+ic)}\, ,\ \ K_2(v+i\gamma)\rightarrow
-\frac{\epsilon}{\delta}\frac{1}{2\pi }\frac{c}{k(k-ic)}\, .  \ee
Introducing $a_1(k)=\au(\delta k/\epsilon-i\g/2),$ $a_2(k)=\ad(\delta
k/\epsilon+i\g/2)$ it is easy now to see that in the continuum limit
(\ref{i15}) transform in the NLIEs (\ref{NLIEgas}) of the Gaudin-Yang model.
The expression for the grandcanonical potential (\ref{Ggas}) is derived from
(\ref{i16}) using $\phi(\mu,H,\bar
\beta)=(f(h_1,h_2,h_3,\beta)-E_0/L)/\delta^3$ with
$f(h_1,h_2,h_3,\beta)=-\ln\Lambda_0(0)/\beta$ and that $K_1(\delta k/\epsilon
-i\g/2)= K_2(\delta k/\epsilon +i\g/2)\sim\epsilon$ in the scaling limit (the
$\ln(1+\mathfrak{a}_2(0))/\delta$ term vanishes in the same limit due to the
fact that the real part of the driving term of Eq.~(\ref{nlie2}) is large and
negative like $\mathcal{O}(1/\epsilon^2)$) .

\section{Conclusions}\label{s10}

We have introduced an efficient thermodynamic description of the repulsive
Gaudin-Yang model which was derived using the connection with the Perk-Schultz
spin chain and the quantum transfer matrix method. Our system of NLIEs is
valid for all values of coupling strengths, chemical potentials and magnetic
fields and can be easily implemented numerically. The numerical data presented
for various thermodynamic quantities reveals the complex interplay between
interaction strength, statistical interaction and temperature. The
nonmonotonicity of the contact as a function of the interaction strength and
temperature shows that the momentum distribution of the repulsive
Gaudin-Yang model has a nontrivial behavior as a function of these parameters
which can be experimentally detected. Compared with the zero temperature case
the density profiles of the trapped system present a double shell structure
only above a critical polarization which depends on the coupling strength and
temperature.  Our paper also opens further avenues of research. A natural
expectation is that the attractive case can also be investigated along similar
lines. An appropriate lattice embedding would also be the Perk-Schultz spin
chain with the same grading but considered in the massive regime. It is also possible
that difficulties can occur due to the presence of additional bound states in
the spectrum.
One could also determine the correlation lengths of the Green's
functions following \cite{KP3} which would require an analysis of the next
leading eigenvalues of the QTM. This will be deferred to future publications.

\section{Acknowledgment}
O.I.P. work was supported from the PNII-RU-TE-2012-3-0196 grant of the
Romanian National Authority for Scientific Research.

\appendix

\section{Solution of the generalized $q=3$ Perk-Schultz model}\label{a1}

In this appendix we present the solution of the generalized $q=3$ Perk-Schultz
model (see the Supplemental Material of \cite{KP2}) which contains as
particular cases the transfer matrix and quantum transfer matrix of the
Perk-Schultz spin chain.  The generalized model is constructed from the
trigonometric Perk-Schultz $\R$-matrix \cite{dV,dVL} defined by
\be\label{rm}
\R(v,w)=\sum_{a=1}^3\R_{aa}^{aa}(v,w)e_{aa}\, \otimes e_{aa}+
\sum_{\substack{a,b=1\\a\ne b}}^3\R_{ab}^{ab}(v,w)\, e_{aa}\otimes e_{bb}
+\sum_{\substack{a,b=1\\a\ne b}}^3 \R_{ba}^{ab}(v,w)\, e_{ab}\otimes e_{ba}\, ,
\ee
with
\be
\R_{aa}^{aa}(v,w)=\frac{\sin[\g+\varepsilon_a(v-w)]}{\sin\g}\, , \ \ \ \R_{ab}^{ab}(v,w)\underset{a\ne b}{=}\frac{\sin(v-w)}{\sin \g}\, ,
\ \ \ \R_{ba}^{ab}(v,w)\underset{a\ne b}{=}e^{isgn(a-b)(v-w)}\, ,
\ee
where $\gamma\in[0,\pi]$ is the anisotropy,
$(\varepsilon_1,\varepsilon_2,\varepsilon_3)$ is the grading
$(\varepsilon_i\in\{\pm 1\})$ and $(e_{ab})_{ij}=\delta_{ia}\delta_{jb}$ is
the canonical basis in the space of $3$-by-$3$ matrices. The generalized model
also depends on three functions $\varphi_i,\, i\in\{1,2,3\}$ which we will
call the parameters of the model. For the transfer matrix the parameters are
$\varphi_1(v)=\alpha_1(v,0)^L\, , \varphi_2(v)=\beta(v,0)^L\,
,\varphi_3(v)=\beta(v,0)^L$ with $L$ the number of lattice sites of the spin
chain while for the quantum transfer matrix $\varphi_1(v)=e^{\beta
  h_1}(\alpha_1(v,-u) \beta(u,v))^{N/2}\, ,$ $\varphi_2(v)=e^{\beta
  h_2}(\beta(v,-u)\beta(u,v))^{N/2}\, ,$ $ \varphi_3(v)=e^{\beta
  h_3}(\beta(v,-u)\alpha_3(u,v))^{N/2}$ with
$u=-J\sin(\epsilon_1\gamma)\beta/N$ where $N$ is the Trotter number and
$h_1,h_2,h_3$ are chemical potentials.  In the above definitions we have
introduced the notations
\[
\alpha_i(v,w)=\frac{\sin[\g+\epsilon_i(v-w)]}{\sin\g}\, ,\ \ \ \beta(v,w)=\frac{\sin(v-w)}{\sin \g}\, .
\]
The generalized model can be solved using the nested Bethe ansatz
\cite{KP2,G1,GS,ACDFR,BR}. The eigenvalues are
$(g_i(v,w)=\alpha_i(v,w)/\beta(v,w))$
\be\label{eigenvalueGM}
\Lambda(v)=\varphi_1(v)\prod_{j=1}^n g_1(v_j^{(1)},v)+\varphi_2(v)\prod_{j=1}^ng_2(v,v_j^{(1)})\prod_{k=1}^mg_2(v_j^{(2)},v)
+\varphi_3(v)\prod_{j=1}^ng_3(v,v_j^{(2)})\, ,
\ee
with $\{v_j^{(1)}\}_{j=1}^n\, ,\{v_k^{(2)}\}_{k=1}^m\, $ satisfying the Bethe ansatz equations
\begin{subequations}\label{BAEGM}
\begin{align}
\frac{\varphi_1(v_s^{(1)})}{\varphi_2(v_s^{(1)})}&=\prod_{\substack{j=1\\j\ne s}}^n\frac{g_2(v_s^{(1)},v_j^{(1)})}{g_1(v_j^{(1)},v_s^{(1)})}
     \prod_{k=1}^m g_2(v_k^{(2)},v_s^{(1)})\, , \  s=1,\cdots,n\, ,\\
\frac{\varphi_2(v_l^{(2)})}{\varphi_3(v_l^{(2)})}&=\prod_{\substack{j=1\\j\ne s}}^m\frac{g_3(v_l^{(2)},v_j^{(2)})}{g_2(v_j^{(2)},v_l^{(2)})}
     \prod_{k=1}^m \frac{1}{g_2(v_l^{(2)},v_k^{(1)})}\, , \  l=1,\cdots,m\, .
\end{align}
\end{subequations}
In this paper the Hamiltonian of the Perk-Schultz spin chain is given by
($\tr(v)$ is the transfer matrix)
\be\label{HPS}
\mathcal{H}_{PS}=J\sin(\epsilon_1 \g)\,  \tr^{-1}(0)\tr'(0)-
\sum_{j=1}^L\sum_{a=1}^3 h_a e_{aa}^{(j)}\, .
\ee
where the second term in the r.h.s of (\ref{HPS}) is a chemical potential term
which commutes with the main component of the Hamiltonian. The energy spectrum
of the Perk-Schultz spin chain is obtained using (\ref{HPS}),
(\ref{eigenvalueGM}) and the fact that the contribution of the chemical
potential term is $h_1(L-n)+h_2(n-m)+h_3m$ \cite{dV}.  In the case of the
quantum transfer matrix it is preferable to work with a different pseudovacuum
(see the Supplemental Material of \cite{KP2}). This change of the pseudovacuum
means that if in the Hamiltonian we consider the grading
$(\epsilon_1,\epsilon_2,\epsilon_3)$ and chemical potentials $(h_1,h_2,h_3)$
in the results (\ref{eigenvalueGM}) and (\ref{BAEGM}) for the quantum transfer
matrix we have to perform a cyclical permutation in the grading and chemical
potentials i.e., $(\varepsilon_1,\varepsilon_2,\varepsilon_3)\rightarrow$
$(\varepsilon_3,\varepsilon_1,\varepsilon_2)$ and $( h_1, h_2,
h_3)\rightarrow(h_3,h_1,h_2)$.

\section{Some useful identities}\label{a2}

Here we present several identities which play an important role in the
derivation of Section \ref{s82}. The first identity is
\be\label{id1}
\lam_1(v)+\lam_2(v)=c \  \frac{\phi_+(v)}{q_2(v)}q_1(v+i\g)q_1^{(h)}(v)\, , \ \ q_1^{(h)}(v)=\prod_{j=1}^{N/2}\sinh(v-{v'}_j^{(1)})\, ,
\ee
with $c$ a constant.  The proof is relatively straightforward. From the
definition (\ref{i3}) we have
\be\label{ii7}
\lam_1(v)+\lam_2(v)=\phi_+(v) \frac{q_1(v+i\g)}{q_1(v)}\frac{p_1(v)}{q_2(v)}\, ,\ \
p_1(v)=\phi_-(v-i\g)q_2(v)e^{\beta h_3}+\phi_-(v)q_2(v-i\g)e^{\beta h_1}.
\ee
The equation $p_1(v)=0$ is equivalent with $\au(v)=-1$ which shows that the
zeros of $p_1(v)$ are the $N/2$ Bethe roots $\{v_{j}^{(1)}\}_{j=1}^{N/2}$ and
$N/2$ holes $\{v_{j}'^{(1)}\}_{j=1}^{N/2}.$ In addition $p_1(v)$ is
quasiperiodic $p_1(v+i\pi)=(-1)^N p_1(v)$ and $\lim_{v\rightarrow\infty}$ $
p_1(v)/(\sinh(v))^N$$=c$. Therefore,
$p_1(v)=c\prod_{j=1}^{N/2}\sinh(v-v^{(1)}_j)
\prod_{j=1}^{N/2}\sinh(v-v'^{(2)}_j)= c\ q_1(v)q^{(h)}_1(v)\, ,$ which
together with (\ref{ii7}) proves (\ref{id1}). In a similar fashion we can show
that
\be\label{id1bis}
\lam_2(v)+\lam_3(v)=c \  \frac{\phi_-(v)}{q_1(v)}q_2(v-i\g)q_2^{(h)}(v)\, , \ \ q_2^{(h)}(v)=\prod_{j=1}^{N/2}\sinh(v-{v'}_j^{(2)})\, ,
\ee
with $c$ a constant. An immediate consequence of (\ref{id1}) and
(\ref{id1bis}) is that for arbitrary $v$ we have
\begin{align}
-\ln\phi_-(v)+\ln q_1(v)-\ln q_2(v-i\g)+\ln q_1^{(h)}(v)-\ln(1+\mathfrak{a}_1(v))+c_1=0\, ,\label{newi}\\
-\ln\phi_+(v)+\ln q_2(v)-\ln q_1(v+i\g)+\ln q_2^{(h)}(v)-\ln(1+\mathfrak{a}_2(v))+c_2=0\, ,\label{newi2}
\end{align}
with $c_{1,2}$ constants.  Another useful identity is
\be\label{id2}
\int_{\mathcal{C}+\mathcal{C}'}d(v-w)\frac{\mathfrak{a}_j'(w)}{1+\mathfrak{a}_j(w)}dw=0\, ,\ \ \ \
d(v-w)=\frac{d}{dv}\ln\sinh(v-w)\, ,
\ee
where again we considered $\g\in(0,\pi/2)$ and the contour $\mathcal{C}'$ is
presented in Fig.\ref{LE} (the lower edge of $\mathcal{C}'$ coincides with the
upper edge of $\mathcal{C}$ but it has opposite orientation).  The proof of
(\ref{id2}) is identical with the one presented \cite{KP2} and will be
omitted.


\begin{thebibliography}{99}
\bibitem{uni}      Introducing a length scale $l_0$ via $c=m g l_0/\hbar^2,$ the units of grandcanonical
                   potential per length and temperature are $\phi_0=\hbar^2/(2 m l_0^3 )$ and $T_0=\hbar^2/(2 m l_0^2 k_B).$
\end{thebibliography}
\end{document}